\documentclass[1p]{elsarticle}
\usepackage{amsmath}
\usepackage{amsfonts}
\usepackage{amssymb}
\usepackage{graphicx}
\usepackage[center]{subfigure}
\usepackage{hyperref}
\usepackage{braket}

\begin{document}

\begin{frontmatter}
\title{Tan relations in one dimension}
\author{Marcus Barth}
\author{Wilhelm Zwerger}
\address{Physik Department, Technische Universit\"at M\"unchen, 
James-Franck-Strasse, D-85748 Garching, Germany}

\begin{abstract}
We derive exact relations that connect the universal $C/k^4$-decay of the momentum 
distribution at large $k$ with both thermodynamic properties and correlation 
functions of two-component Fermi gases in one dimension with contact interactions. 
The relations are analogous to those obtained by Tan in the 
three-dimensional case and are derived from an operator product expansion of 
the one- and two-particle density matrix. They extend earlier results by
Olshanii and Dunjko~\cite{Olshanii_2003_1D} for the bosonic Lieb-Liniger gas. 
As an application, we calculate the pair distribution function at short distances
and the dimensionless contact in the limit of infinite repulsion. The ground state energy 
approaches a universal constant in this limit, a behavior that also holds in the 
three-dimensional case. In both one and three dimensions, a Stoner instability to 
a saturated ferromagnet for repulsive fermions with zero range interactions is ruled out
at any finite coupling.
\end{abstract}

\begin{keyword}
1D Fermi gas \sep contact interaction \sep OPE \sep ferromagnetism \\
PACS:\,03.75.Ss, 67.10.-j, 67.85.-d
\end{keyword}

\end{frontmatter}

\section{Introduction} \label{sec:introduction}
The study of a non-relativistic system of fermions with spin-independent two-body interactions appears as a 
generic many-body problem in different areas of physics. Except for the particular case of one dimension (1D) 
~\cite{Lieb_Mattis, Sutherland, Giamarchi}, there are, unfortunately, very few exact results on this problem 
beyond the perturbative regime, which is typically not the one that is realized in nature. It is therefore of considerable
interest to derive relations for the many-body problem that hold independent of the interaction strength. A well known 
example is Landau's Fermi liquid theory, which provides exact results for the low-energy excitations and the resulting 
response functions in terms of a few phenomenological parameters. The validity of these relations, however, requires 
that the interacting ground state is adiabatically connected to that of free fermions, 
an assumption that can hardly ever be justified starting from the microscopic Hamiltonian.  Remarkably, considerable 
progress and insight into the fermionic many-body problem at arbitrary strength of the interactions has been achieved 
recently by Sinha Tan~\cite{Tan_2008_energetics} in the particular case 
of zero range interactions. In this special case, it turns out that the momentum distribution exhibits a universal $C/k^4$ 
decay as $k$ approaches infinity. The constant $C$ is the same for both particle species~\cite{Tan_2008_energetics} 
and is called the \textit{contact}, because it is a measure of the probability that two fermions with opposite spin 
are close together~\cite{Tan_2008_energetics} (for a comprehensive recent review of the Tan relations see 
~\cite{Braaten_2010_Book}). The contact determines  the change of the energy with respect to the interaction strength by 
a Hellman-Feynman like relation, the Tan adiabatic theorem~\cite{Tan_2008_adiabatic}. It also allows to calculate the  
energy from the momentum distribution~\cite{Tan_2008_energetics}.  A crucial feature of the Tan relations is the fact that 
they  apply to {\it any} state of the system, e.g. both to a Fermi liquid or a superfluid state, at zero or at finite temperature 
and also in a few-body situation.  The only change is the value of the contact $C$. The origin of this universality was elucidated 
by Braaten and Platter~\cite{Braaten_2008_short} who have shown that the Tan relations are a consequence of operator 
identities that follow from a Wilson operator product expansion of the one-particle density matrix. 

Our aim in the present work is to derive the analog of the Tan relations for two-component Fermi systems in 1D. At first sight, 
this appears to be of little interest because 1D fermions with zero range interactions can be solved exactly by the Bethe-Ansatz
~\cite{Gaudin_1967, Yang_1967, Batchelor_2006_Bethe}.
As will be shown below, however, the Tan relations in 1D provide information on observables that are not easily calculable from 
the Bethe-Ansatz, like the momentum or the pair distribution function. Moreover, they also apply at finite temperatures or in the
presence of an external confining potential, where the Bethe-Ansatz fails. 
This is of particular relevance for the case of ultracold fermionic atoms, for which a model with contact interactions in fact provides 
a realistic description of atoms confined into a 1D quantum wire geometry~\cite{Bloch_2008_review}. In particular, 
the use of Feshbach resonances in this context allows to tune the interaction strength simply by changing a magnetic field.
This gives access to the whole regime from weak interactions to the limit of  infinite attraction or repulsion. In the 3D case, the latter limit 
is reached in the unitary Fermi gas, where the two-body scattering length $a$ diverges.  The attractive branch of the unitary Fermi gas has
a superfluid ground state and has been studied quite extensively in the context of the BCS--BEC crossover problem 
~\cite{Ketterle_2008_review, Bloch_2008_review, Giorgini_2008_review}.  It provides an example of a non-relativistic field theory that 
is both scale and conformally invariant at $k_Fa=\infty$~\cite{Son_2006_symmetry, Nishida_2007_CFT}. The nature of the ground state 
on the - metastable - repulsive branch, in turn, is still unknown beyond the perturbative limit which -- by construction -- is a Fermi liquid  
~\cite{Abrikosov_1963}. In particular, it is an open question whether the repulsive 3D gas exhibits a Stoner type instability to a
ferromagnetic state, as indicated by a renormalized Hartree-Fock calculation~\cite{Duine_2005}. Experimental support for the existence 
of such an instability has been provided by the observation of a sudden decrease in the three-body loss rate and a minimum in the kinetic 
energy of the gas at a critical value $k_Fa=1.9\pm 0.1$ of the interaction parameter~\cite{Jo_2009_ferro}. Recent variational 
~\cite{Zhai_10} and numerical calculations~\cite{Bertaini_2010_ferro, Trivedi_2010_ferro} in fact find that the ground state energy 
of the unpolarized Fermi liquid state exceeds that of a {\it saturated} ferromagnet for sufficiently strong repulsion. The associated value 
of the critical coupling constant turns out to be much smaller than that observed experimentally, a discrepancy that might be explained by the fact that the 
repulsive branch is only metastable.  By contrast, as will be shown below, even in an equilibrium situation where the lifetime of the repulsive 
branch is assumed to be infinite, a combination of the Tan relations and a variational argument rules out a Stoner instability to a 
saturated ferromagnet for fermions with zero range interactions in both one and in three dimensions.

Our derivation of the Tan relations is based on the Operator Product Expansion (OPE), which was developed independently by 
Wilson~\cite{Wilson_1969} and Kadanoff~\cite{Kadanoff_1969}. In fact, this method was implicitly used in this context first by 
Olshanii and Dunjko~\cite{Olshanii_2003_1D} in a study of the short distance behavior of the one-particle density matrix of the Lieb-Liniger 
gas, a system of 1D bosons with zero range interactions~\cite{Lieb_Liniger_1963_Bose}. Specifically, they found that the momentum 
distribution of this bosonic system decays like $C_B / k^4$ for large $k$. The associated contact coefficient $C_B$ turned out to be proportional 
to the pair distribution function $g^{(2)}(0)$ at vanishing distance between the particles. As will be shown in section 4 below, this is effectively the 
1D analog of the Tan relation for the asymptotic decay of the momentum distribution for bosons. It carries over to the two-component Fermi gas 
with only slight modifications.
%  \footnote{The difference to ~\cite{Olshanii_2003_1D} is due to a different normalization of the momentum distribution - we normalized it to give 
%the \textit{particle density} when integrated over all momenta.}:
%\begin{equation}
%\label{eq:bosonic-contact}
 %C_B = \gamma^2 n^4 e'(\gamma)= \gamma^2 n^4 g^{(2)}(0)
%\end{equation}
%Here, $n$ is the 1D density and $\gamma=2/n |a_1|>0$ the dimensionless coupling constant which involves the 1D scattering length $a_1$ (for the precise %definition of $a_1$ see section .. below). A very similar relation will be found below also for a two-component Fermi gas. This shows that Bosons and %fermions in 1D are very similar not only in their long-wavelength behavior, which is described by  Luttinger liquid theory ~\cite{Haldane_1981, Giamarchi},  %but also as far as short distance physics is concerned.

The paper is organized as follows: in section 2 we introduce our model and discuss its range of validity. The Tan adiabatic 
theorem is derived in section 3 as a simple application of the Hellman-Feynman theorem. In section 4 the 
OPE of the one-particle density matrix is used to connect the asymptotics of the momentum distribution with the pair distribution
function for opposite spins at vanishing separation, which is essentially the contact in one dimension. 
Thermodynamic relations that connect energy and pressure with the contact are derived in section 5. An OPE of
the density correlation function in section 6 shows that the contact also appears as a non-analytic contribution to
the pair distribution function at short distance, giving rise to sum rules and a power law asymptote in the static structure factor.
Explicit results for the contact at arbitrary interaction strengths are given in section 7 for the particular case of
the balanced Fermi gas at zero temperature on the basis of the known Bethe-Ansatz solution for the ground state
energy. Finally, in section 8, we show that irrespective of an explicit solution of the many-body problem, a combination of the Tan
relations and a variational argument rules out saturated ferromagnetism for zero range interactions at any finite coupling, both in one and also
in three dimensions. Details of the diagrammatic expansion for the OPE are presented in an Appendix.

%%%%%%%%%%%%%%%%%%%%
%Hamiltonian etc.
\section{The 1D Fermi gas with contact interactions} \label{sec:model}
We consider a Gaudin-Yang model for a two-component system of fermions in 1D which interact with a two-body $\delta$-function
potential. Due to the Pauli principle, this interaction only affects fermions with opposite 'spin'  $\sigma =\uparrow,\downarrow$. In the 
case of ultracold atoms,  the two spin states might be two different hyperfine states of  $^6$Li or $^{40}$K or they might describe 
a fermionic mixture with different masses $m_{\sigma}$. In a second quantized form, the Hamiltonian is expressed in terms of 
field operators  $\psi_{\sigma} (R)$ which obey the usual anti-commutation relations.  Including an external one-body potential $\mathcal{V}(R)$ 
and using units in which $\hbar = 1$,  the Hamiltonian $H = \int dR\,\mathcal{H}(R)$ can be written in terms of a Hamiltonian density 
\begin{equation}
 \label{eq:hamilton-density}
 \mathcal{H}(R) = \sum_{\sigma} \left[ \frac{1}{2m_{\sigma}} \partial_R \psi^{\dagger}_{\sigma} \partial_R \psi_{\sigma}(R) 
+
\frac{g_{1}}{2}\psi_{\sigma}^{\dagger}\psi_{-\sigma}^{\dagger}\psi_{-\sigma}\psi_{\sigma}(R)\right]
+ \sum_{\sigma}\mathcal{V}_{\sigma}(R)\psi^{\dagger}_{\sigma}\psi_{\sigma}(R).
\end{equation}
The $\delta$-function interaction is a proper potential in 1D  and thus needs no renormalization. Its coupling constant 
$g_1$ is conveniently expressed in terms of a one-dimensional scattering length $a_1$ by $g_1 = -1/m_r a_1$, where 
$m_r$ is the reduced mass of the two interacting particles. In order to discuss the physical meaning of this scattering 
length and the question under which conditions a realistic two-body interaction potential can be replaced by an 
effective contact interaction, we consider the low-energy limit  
\begin{equation}
r(k)=\frac{-1}{1+i\cot(\delta(k))}\simeq \frac{-1}{1+ika_1+{\cal O}\left((k\ell_{\perp})^3\right)}
\label{eq:reflection}
\end{equation}
of the reflection amplitude that describes two-particle scattering in one dimension quite generally in a situation, where the 
particles are  confined to the ground state of their transverse motion with size $\ell_{\perp}$ ~\cite{Bloch_2008_review}.   
For a gas with total density $n$, the momenta for two-body scattering are 
typically of the order of the Fermi momentum $k_F=\pi n/2$. The reflection amplitude 
can thus be replaced by its low-energy limit $-1/(1+ika_1)$ provided that $k_F^2\ll a_1/\ell_{\perp}^3$.
In a regime of low densities, therefore, any two-body interaction for which the low-energy expansion
in equation (\ref{eq:reflection}) holds, can be replaced by an effective contact interaction $g_1\delta(x)$
with coupling constant $g_1=-1/m_ra_1$, for which the reflection amplitude equals $-1/(1+ika_1)$ at
{\it arbitrary} values of $k$. Specifically, for ultracold atoms whose interaction is described by a 
3D pseudopotential with scattering length $a$, the resulting value of $a_1$ is given by ~\cite{Olshanii_1998}
\begin{equation}
a_1(a)=-\frac{\ell_{\perp}^2}{a}+A\ell_{\perp}\, ,
\label{eq:a1}
\end{equation}
where $A=-\zeta(1/2)/\sqrt{2}\simeq 1.0326$ is a numerical constant. Note, that 
a negative value $a_1<0$ of the 1D scattering length corresponds 
to repulsive interactions and vice versa, a situation that is precisely opposite to the 3D case. For $a_1>0$,
the scattering amplitude has a pole at $k=i/a_1$ with a positive imaginary part. This describes the two-body
bound state of an attractive $\delta$-function potential, with $a_1$ the size of this bound state. 

\begin{figure}
\begin{center}
\begin{minipage}{50pt}
 \includegraphics[width=50pt,height=75pt]{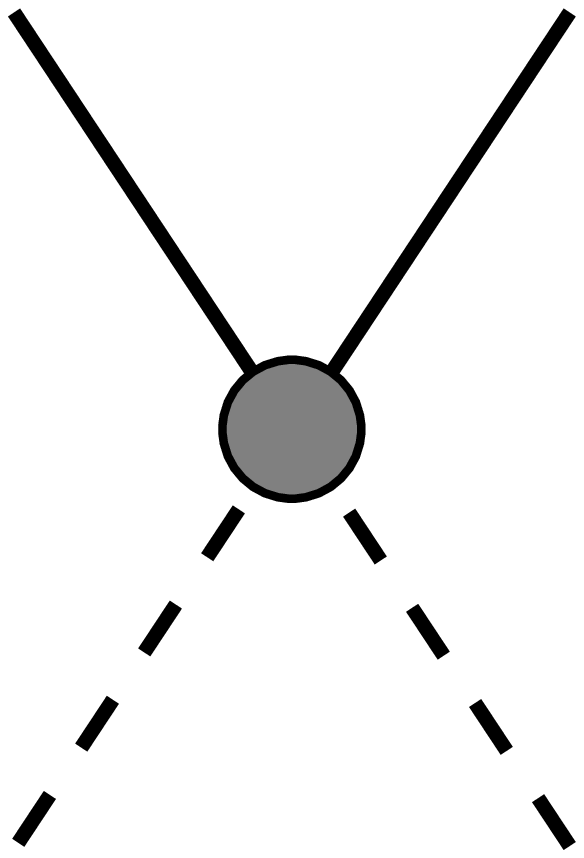}
\end{minipage}
=
\begin{minipage}{50pt}
 \includegraphics[width=50pt,height=75pt]{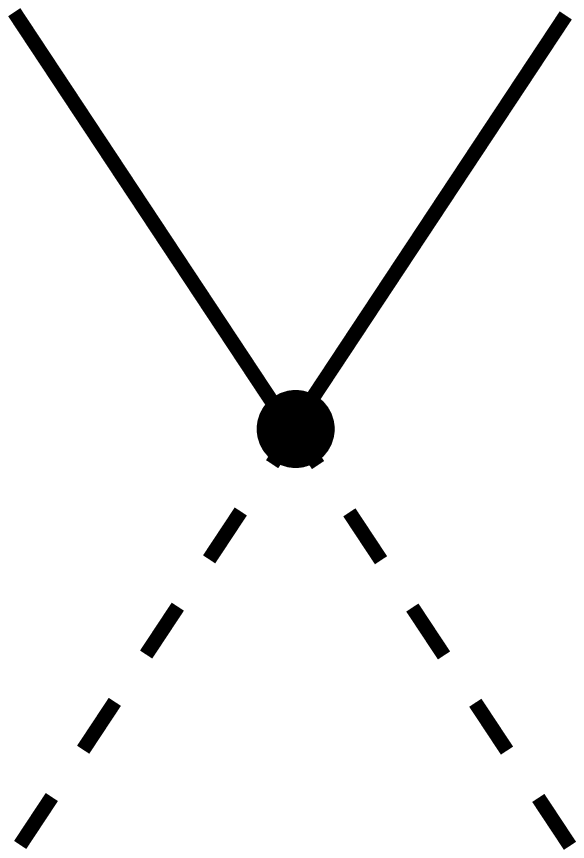}
\end{minipage}
+
\begin{minipage}{111pt}
 \includegraphics[width=111pt,height=75pt]{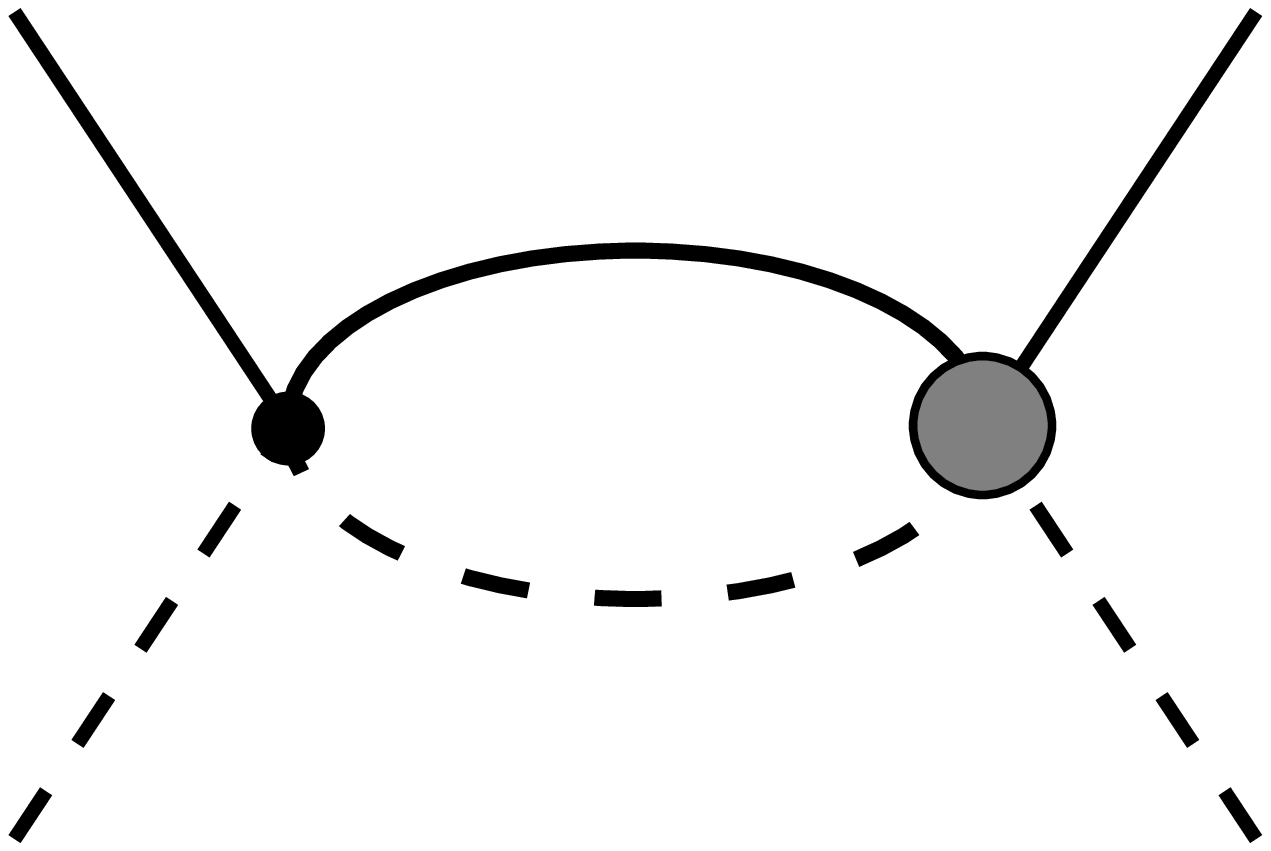}
\end{minipage}
\end{center}
%caption of the Lippman-Schwinger equation figure
\caption{Integral equation for the scattering of a pair of fermions. The big gray blob represents the amplitude $i \mathcal{A}(E)$.}
\label{fig:lippmann-schwinger}
\end{figure}

For the calculation of the diagrams that appear in the Wilson-OPE for the one- and two-body density matrix 
in sections 4 and 6 below, it is convenient to use the amplitude $\mathcal{A}(E)$, which is related to the 
on-shell $T$-Matrix at the center of mass energy $E=k^2/2m_r$ by $\mathcal{A}(E)=-T(E)$.   It obeys the 
Lippmann-Schwinger equation shown diagrammatically in figure \ref{fig:lippmann-schwinger}. 
For the $\delta$-function interaction, the Lippmann-Schwinger equation is solved analytically by
\begin{equation}
 \label{eq:scattering-amplitude}
 \mathcal{A}(E)=\frac{-ik}{m_r}\cdot r(k)=\frac{1}{m_r} \cdot  \frac{1}{a_1-\frac{i}{\sqrt{2m_rE}} }\, 
\end{equation} 
which is just the associated refection amplitude up to a factor $-ik/m_r$. 
   
%%%%%%%%%%%%%%%%%%%%%%%%%%%%%%%%%%%%%%%%%%
%begin: Lippmann-Schwinger equation figure
%end: Lippmann-Schwinger equation figure
%%%%%%%%%%%%%%%%%%%%%%%%%%%%%%%%%%%%%%%%

\section{The Tan adiabatic Theorem} \label{sec:adiabatic}
For 1D systems with zero range interactions, the expectation value of the interaction part 
of the Hamiltonian
\begin{equation}
 \langle H'\rangle=g_1 \int dR \langle \psi^{\dagger}_{\uparrow}\psi^{\dagger}_{\downarrow}\psi_{\downarrow}\psi_{\uparrow}(R) \rangle
 \label{eq:Hprime}
\end{equation} 
is quite generally determined by the value of the pair distribution function for opposite spins
\begin{equation}
g_{\uparrow\downarrow}^{(2)}(R,R')  =  \langle \psi^{\dagger}_{\uparrow}(R)\psi^{\dagger}_{\downarrow}(R')\psi_{\downarrow}(R')\psi_{\uparrow}(R) \rangle
/\left(n_{\uparrow}(R)n_{\downarrow}(R')\right)
 \label{eq:gupdown}
\end{equation} 
at vanishing distance $R=R'$ between the particles, as noted by Lieb and Liniger
in the Bose gas case ~\cite{Lieb_Liniger_1963_Bose}.  The interaction energy in \eqref{eq:Hprime} is therefore 
simply $g_1$ times an integral of the probability density for two fermions with opposite spin to be at the same point 
in space,  a quantity that is proportional to the contact density discussed in section 4 below. More precisely, 
it is convenient to define the contact $C$ as an extensive variable by 
\begin{equation}
 \label{eq:contact-exact}
 C=4g_1m_r^2 \langle H'\rangle=4 g_1^2 m_r^2 \int dR\, 
 \langle \psi^{\dagger}_{\uparrow}\psi^{\dagger}_{\downarrow}\psi_{\downarrow}\psi_{\uparrow}(R) \rangle.
\end{equation}
This variable will play a central role in the following. It depends both on the microscopic parameters like 
the scattering length $a_1$ and also on the specific state under consideration. While the contact is basically the
expectation value of the interaction part of the Hamiltonian, it also determines the variation of the total 
energy $E$ with the scattering length $a_1$. Indeed, by the Hellman-Feynman theorem, we have  
\begin{equation}
 \frac{dE}{da_1} = \Braket{ \frac{\partial H}{\partial a_1} } = \frac{1}{a_1^2} \frac{1}{m_r} 
  \int dR\, \langle \psi^{\dagger}_{\uparrow}\psi^{\dagger}_{\downarrow}\psi_{\downarrow}\psi_{\uparrow}(R) \rangle\, .
\end{equation}
Using the definition of the contact, this immediately gives the 1D analog of the Tan adiabatic theorem  
~\cite{Tan_2008_adiabatic} in the form
\begin{equation}
 \label{eq:adiabatic-relation}
 \frac{d}{da_1} E = \frac{C(a_1)}{4m_r}.
\end{equation}
A knowledge of the dependence of the contact $C$ on the scattering length therefore determines the associated rate of change 
of the total energy.  Note that the derivative in \eqref{eq:adiabatic-relation} has to be taken with all other variables constant, 
in particular the entropy if thermal states are considered.  This will play a crucial role in the derivation of the pressure relation
in section 5. A particularly simple form of the Tan adiabatic theorem is obtained by considering a translation invariant situation with
vanishing external potential $\mathcal{V}\equiv 0$. Applying the Hellman-Feynman theorem to the Hamiltonian density in equation 
(\ref{eq:hamilton-density}) then shows that the pair distribution function $g_{\uparrow\downarrow}^{(2)}(0)$  for opposite 
spins at zero distance is determined by the derivative of the energy per length with respect to the 
coupling constant $g_1$ via
\begin{equation}
   \frac{\partial\langle\mathcal{H}\rangle}{\partial g_1}=n_{\uparrow}n_{\downarrow}\, g_{\uparrow\downarrow}^{(2)}(0) \, .
 \label{eq:Hellman-Feynman}
\end{equation}
This relation will be used in section 7 below to determine the contact explicitly in the ground state of the balanced two-component 
Fermi gas from the known Bethe-Ansatz result for the ground state energy for arbitrary coupling strength.   
 
%%%%%%%%%%%%%%%%%%%%%%%%%%%%%%%%%%
%study of the asymptotc behaviour
\section{Asymptotics of the Momentum Distribution} \label{sec:asymptotics}
In the following, we want to show that the contact introduced in \eqref{eq:contact-exact} determines not only 
the interaction energy and the derivative of the total energy with respect to $a_1$ but also appears in the
asymptotic behavior of the momentum distribution 
\begin{equation}
\label{eq:ntilde-def}
 \tilde{n}_{\sigma}(k) = \langle \tilde{\psi}^{\dagger}_{\sigma} (k) \tilde{\psi}_{\sigma} (k) \rangle,
\end{equation}
where $\tilde{\psi}_{\sigma}(k)$ is the Fourier transform of the field $\psi_{\sigma}$. The momentum distribution
$\tilde{n}_{\sigma}(k)$ is normalized such that the integral over all momenta gives the total number $N_{\sigma}$ 
of particles of type $\sigma$. This will be more convenient for the following arguments compared to the standard 
intensive normalization of $n_{\sigma}(k)$ to the respective densities $n_{\sigma}$. 
In terms of the fields $\psi_{\sigma}^{\dagger}(R)$ and $\psi_{\sigma}(R)$, the momentum distribution can 
be expressed by
\begin{equation}
\label{eq:ntilde-1pd-matrix}
 \tilde{n}_{\sigma} (k) = \int dR \int dx \, e^{-ikx} \langle \psi^{\dagger}_{\sigma} (R) \psi_{\sigma} (R+x) \rangle \, .
\end{equation}
Its behavior at large momenta $k$ can therefore be obtained from an expansion of the one-particle density matrix  
$\langle{\psi}_{\sigma}^{\dagger}(R) {\psi}_{\sigma}(R+x)\rangle$ for small separations $x\to 0$. To do this
in practice, we use the Wilson-Kadanoff Operator Product Expansion (OPE) which states that the product of two 
local operators separated by a short distance, 
can be expanded as the sum of local operators with coefficients that depend only on the separation. 
For the specific case of the one-particle density matrix, this statement reads
\begin{equation}
\label{eq:OPE-1pdm-general}
 \psi^{\dagger}_{\sigma} (R) \psi_{\sigma} (R+x) = \sum_{n} c_{\sigma , n}(x) \mathcal{O}_{\sigma,n}(R),
\end{equation}
where the local operators $\mathcal{O}_{\sigma,n}(R)$ are products of the quantum fields and their derivatives. The Wilson coefficients $c_{\sigma,n}(x)$ encode all the short distance behavior, in particular possible singularities. The expansion is to be understood as an asymptotic one, i.e. for any positive integer $l$ only a finite number of terms on the right hand side vanish more slowly than $|x|^l$ as $x \rightarrow 0$.  Knowledge of the leading terms in the OPE determines the large $k$ behavior of the momentum distribution via
\begin{equation}
\label{eq:momentumdistribution-OPE}
 \tilde{n}_{\sigma}(k) = \sum_n  \int dx \, e^{-ikx} c_{\sigma ,n}(x) \int dR \langle \mathcal{O}_{\sigma,n}(R) \rangle .
\end{equation}

%\paragraph*{Matching}
Since \eqref{eq:OPE-1pdm-general} is an operator equation, it holds for matrix elements between arbitrary states. This allows to determine the Wilson coefficients $c_{\sigma ,n}(x)$ by a matching procedure. We compute the matrix elements of the left hand side and right hand side of equation \eqref{eq:OPE-1pdm-general} between states $\bra{\chi}, \ket{\phi}$, expand the left hand side with respect to $x$, and demand  both sides of the equation to be equal up to some order in $x$. The Wilson coefficient of an operator $\mathcal{O}_{\sigma,n}(R)$ may be determined from the matrix element between any states, for which the matrix element $\bra{\chi} \mathcal{O}_{\sigma,n}(R) \ket{\phi}$ is non-zero. Except for this constraint, one may choose the simplest possible states one can think of. \\
Our explicit calculations follow closely the method used by Braaten and Platter in the three-dimensional case ~\cite{Braaten_2008_short, Braaten_2008_long}. For the coefficients of one-particle operators such as $\psi^{\dagger}_{\sigma} \psi_{\sigma}$, we will choose one-particle
plane wave states $\bra{p'}, \ket{p}$ containing one particle of the species $\sigma$ with momentum $p'$ or $p$ respectively. For two-particle operators, such as $\psi_{\uparrow}^{\dagger} \psi_{\downarrow}^{\dagger} \psi_{\downarrow} \psi_{\uparrow}$, we choose two-particle \textit{scattering states} $\bra{\pm p'}$, $\ket{\pm p}$ in the center of mass frame, containing one particle of each species with momenta $\pm p$ and $\pm p'$ respectively. The computation of these matrix elements will be simplified by drawing the corresponding diagrams and using the Feynman rules for the theory and the operator vertices given in \ref{app:feynman}.
%%%%%%%%%%%%%%%%%%%%%%%%%%%%%%%%%%%%%%%%%%%%%%%%%%%%%%%%%%%%%%%%%%%%%%%%%%%%
%begin: one-particle diagrams for the matching of the one-particle operators
\begin{figure}
\begin{center}
\subfigure[{}]{\label{subfig:1plocal_lhs}\includegraphics[width=150pt,height=30pt]{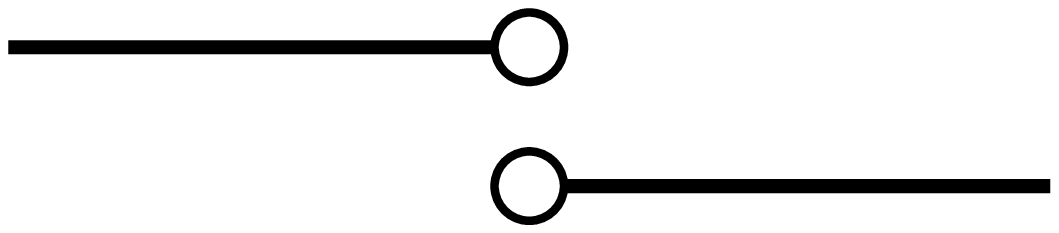}}
\subfigure[{}]{\label{subfig:1plocal_rhs}\includegraphics[width=150pt,height=11pt]{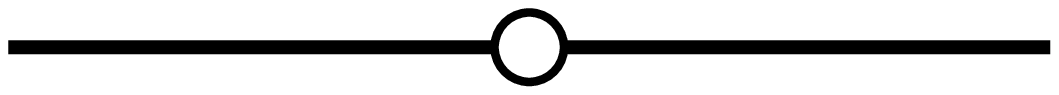}}
\end{center}
%caption one-particle diagrams
\caption{Diagrams for the matrix elements of one-particle operators between one-particle states.}
\label{fig:localbilocalop-matrixelements-1pstate}
\end{figure}
%end: one-particle diagrams for the matching of the one-particle operators
%%%%%%%%%%%%%%%%%%%%%%%%%%%%%%%%%%%%%%%%%%%%%%%%%%%%%%%%%%%%%%%%%%%%%%%%%%
We will consider the OPE of $\psi_{\uparrow}^{\dagger}(R) \psi_{\uparrow} (R+x)$ to be able to draw the diagrams in a definite way -- the matching for the one-particle density matrix of species $\sigma=\downarrow$ is completely analogous. The matrix elements for the matching of the one-particle operators can be represented by the diagrams shown in figure \ref{fig:localbilocalop-matrixelements-1pstate}. In particular, the matrix element of the 
one-body density matrix in single-particle plane wave states is equal to
\begin{equation}
 \label{eq:1p-matching-lhs}
 \bra{p'} \psi_{\uparrow}^{\dagger}(R) \psi_{\uparrow} (R+x) \ket{p} = e^{i(p-p')R} e^{ipx}.
\end{equation}
The right hand side consists of operators of the type $\psi_{\uparrow}^{\dagger} (\partial_R)^m \psi_{\uparrow}$, whose matrix elements are given by
\begin{equation}
 \label{eq:1p-matching-rhs}
 \bra{p'} \psi_{\uparrow}^{\dagger} (\partial_R^m \psi_{\uparrow}) (R) \ket{p} = (ip)^m e^{i(p-p')R}.
\end{equation}
Comparing both sides of the OPE, the Wilson coefficients of the local one-particle operators $\psi_{\uparrow}^{\dagger} (\partial_R)^m \psi_{\uparrow}$ have to match the exponential term $e^{ipx}$ and are thus given by $x^m / m!$. The contribution coming from these operators is analytic in $x$ and is just equivalent to a Taylor expansion of the one-particle density matrix around $x=0$.  Note the slight difference compared to the matching performed in ~\cite{Braaten_2008_short}, because we have put the $x$-dependence only in the argument of the $\psi_{\sigma}$-field. Thus, there are no analytic contributions involving derivatives of $\psi^{\dagger}_{\sigma}$ in our case.
%%%%%%%%%%%%%%%%%%%%%%%%%%%%%%%%%%%%%%%%
%begin: figure bilocal operator diagrams
\begin{figure}

\begin{center}
\subfigure[{ }]{\label{subfig:1pdm1} \includegraphics[width=50pt,height=75pt]{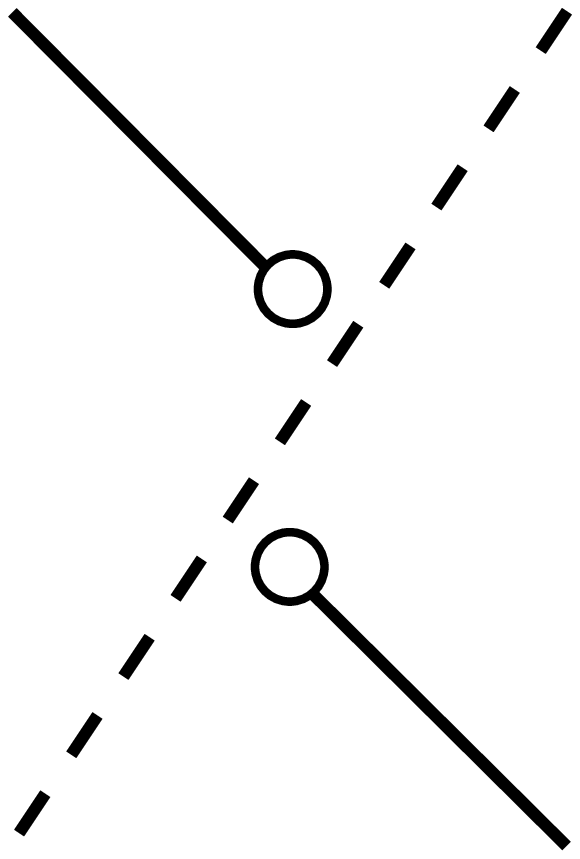}} \hspace{5pt}
\subfigure[{ }]{\label{subfig:1pdm2} \includegraphics[width=50pt,height=75pt]{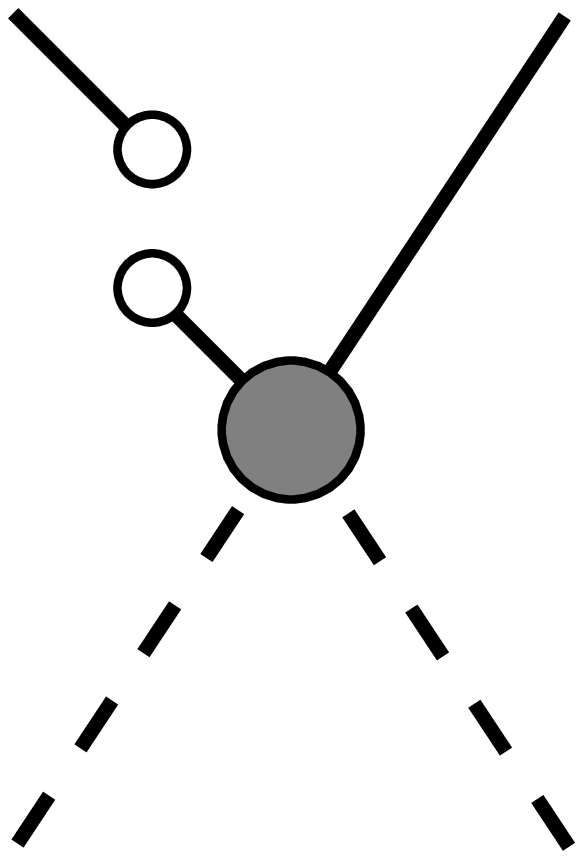}} \hspace{5pt}
\subfigure[{ }]{\label{subfig:1pdm3} \includegraphics[width=50pt,height=75pt]{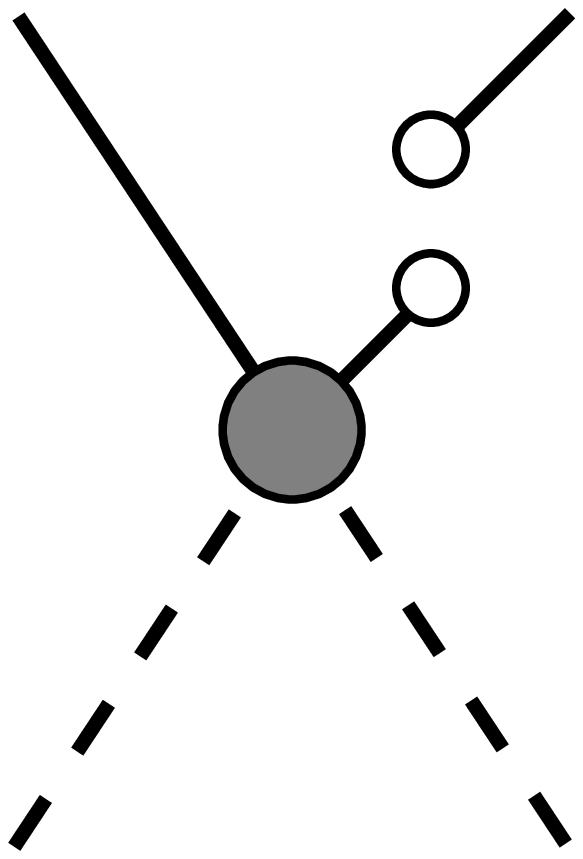}} \hspace{5pt}
\subfigure[{ }]{\label{subfig:1pdm4} \includegraphics[width=111pt,height=75pt]{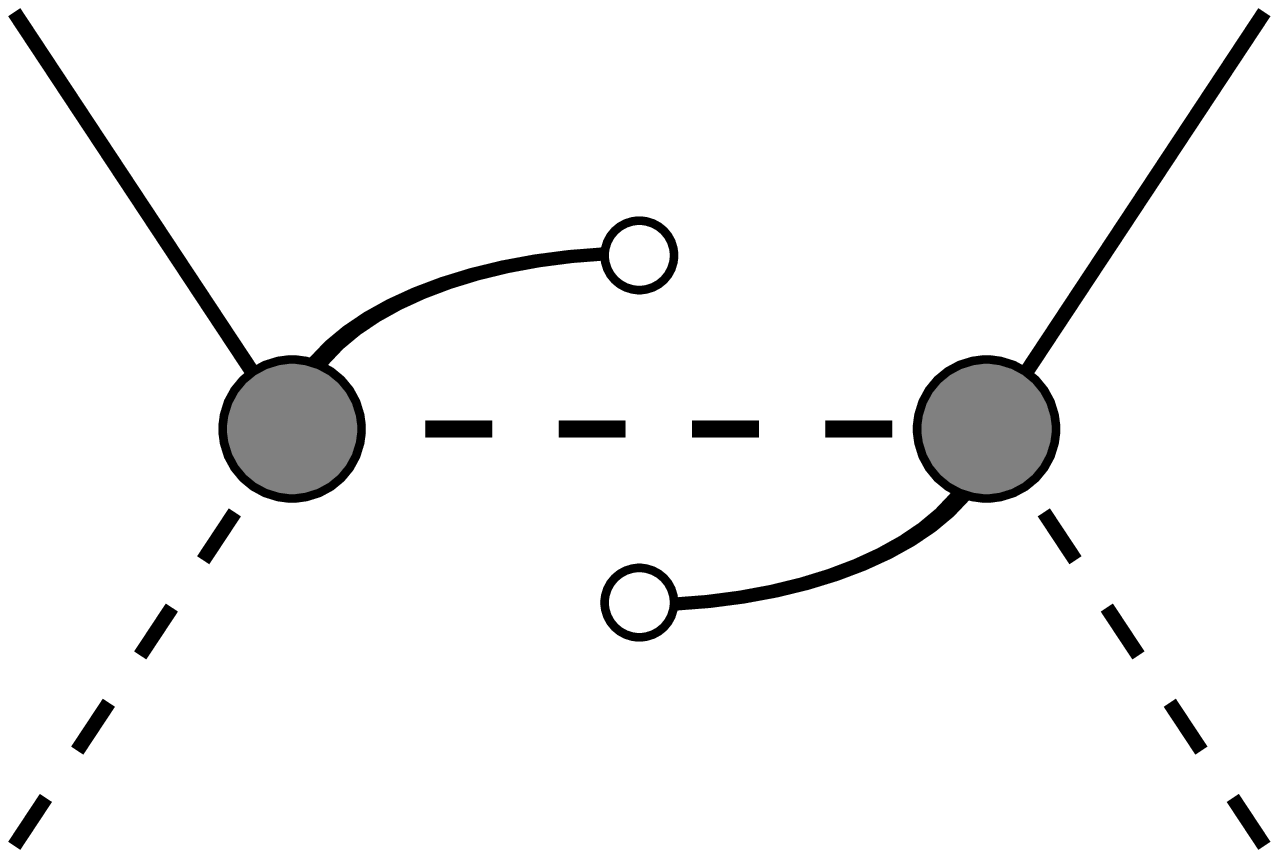}}
\end{center}

%caption: bilocal operator matrix elements
\caption{Diagrammatic contributions to the left hand side of the OPE of the one-particle density matrix, using two-particle scattering states.}
\label{fig:bilocalop-matrixelements}
\end{figure}
%end: figure bilocal operator diagrams
%%%%%%%%%%%%%%%%%%%%%%%%%%%%%%%%%%%%%%
In analogy to the three-dimensional case there are, however, also non-analytic contributions in the short distance expansion of the one-particle
density matrix. The  leading term is, in fact, again connected with the operator 
$\psi_{\uparrow}^{\dagger} \psi_{\downarrow}^{\dagger} \psi_{\downarrow} \psi_{\uparrow}$ that also appears in 3D ~\cite{Braaten_2008_short}. To show this, we calculate the matrix element of the one-particle density matrix between the two-particle scattering states. Diagrammatically, this can be represented by the sum of the four diagrams shown in figure \ref{fig:bilocalop-matrixelements}. The first three diagrams with one or no scatterings contribute purely analytic terms. Together with the analytic part of diagram \ref{subfig:1pdm4}, they are matched by the one-particle operators whose Wilson coefficients are already known. The high momentum region in the loop integral gives rise to a non-analytic contribution to the diagram \ref{subfig:1pdm4} of the form
\begin{equation}
 \label{eq:2p-matching-lhs-nonanalytic}
 4m_r^2 \mathcal{A}(E)\mathcal{A}(E') \frac{i}{2} \frac{1}{|p|^2-|p'|^2} \left[  \frac{e^{i|p||x|}}{|p|} - \frac{e^{i|p'||x|}}{|p'|} \right],
\end{equation}
where $\mathcal{A}(E)$ is the scattering amplitude given in equation \eqref{eq:scattering-amplitude} while $E=p^2/2m_r$ and $E'=p^2/2m_r$
are the associated energies.  The expansion of the bracket in equation \eqref{eq:2p-matching-lhs-nonanalytic} in powers of $|x|$ has a leading order non-analytic term
proportional to $|x|^3$, which, after Fourier transform, will give rise to a $1/k^4$ tail in the momentum distribution.
%%%%%%%%%%%%%%%%%%%%%%%%%%%%%%%
%begin: figure contact diagrams
\begin{figure}

\begin{center}
\subfigure[{ }]{\label{subfig:contact1} \includegraphics[height=75pt]{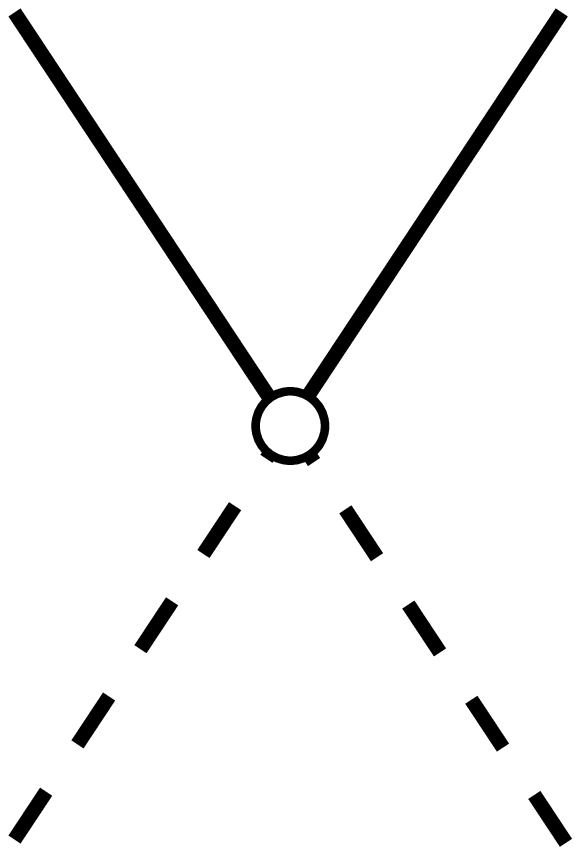}} \hspace{5pt}
\subfigure[{ }]{\label{subfig:contact2} \includegraphics[height=75pt]{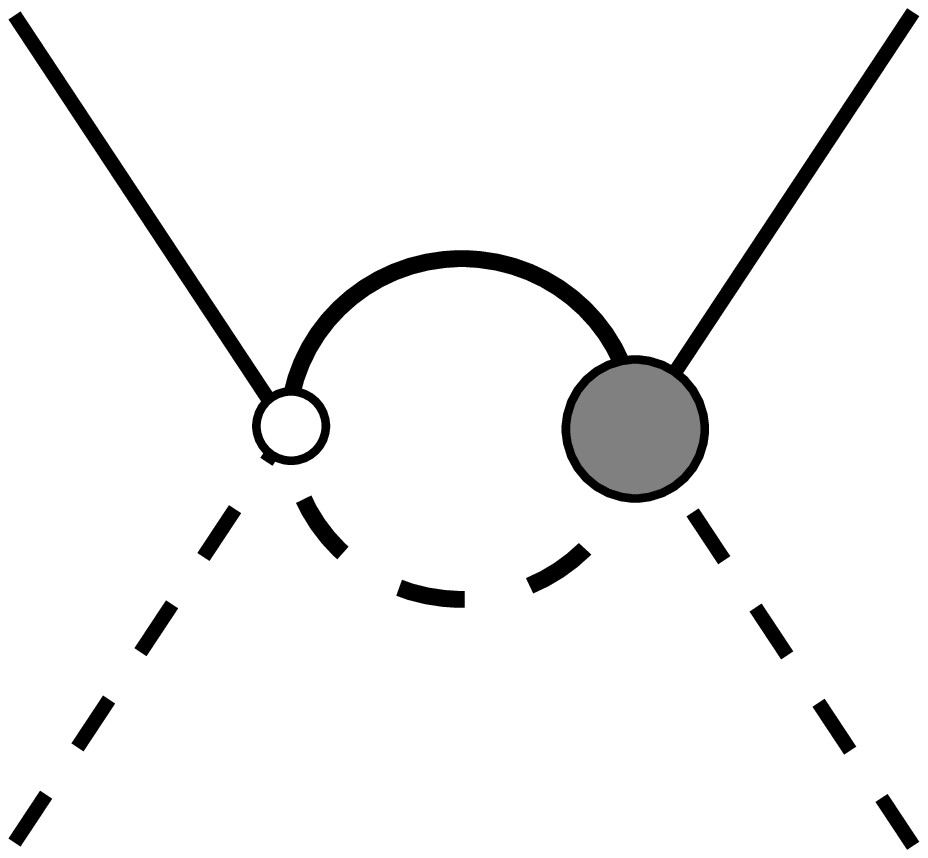}} \hspace{5pt}
\subfigure[{ }]{\label{subfig:contact3} \includegraphics[height=75pt]{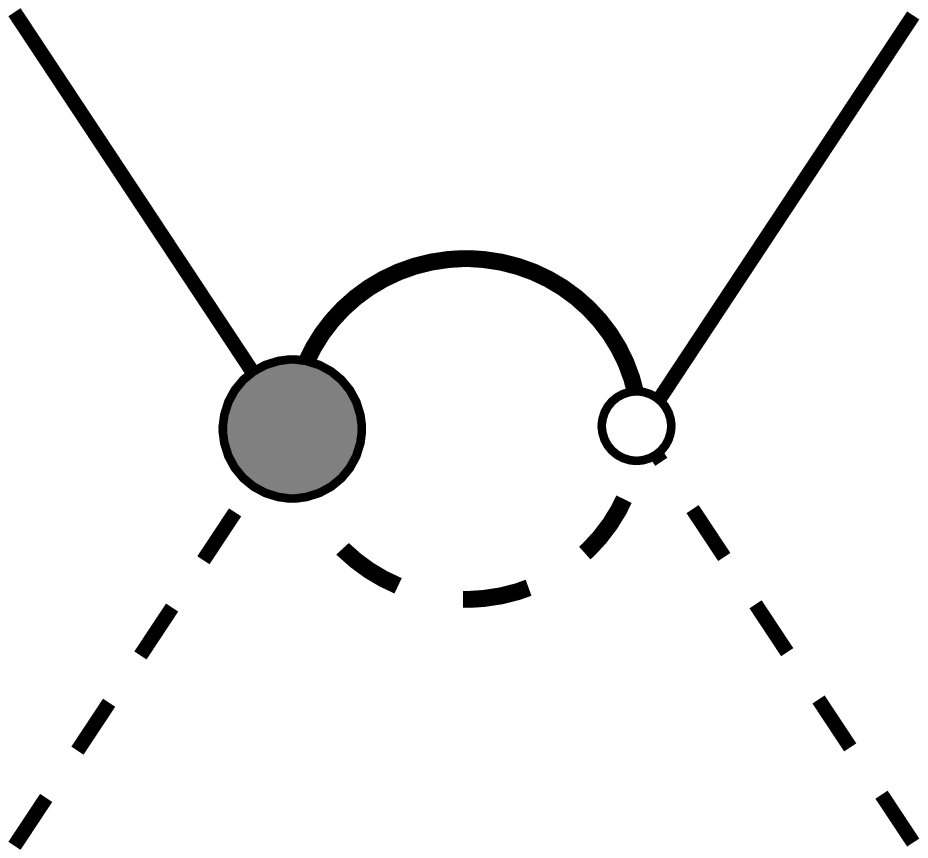}} \hspace{5pt}
\subfigure[{ }]{\label{subfig:contact4} \includegraphics[height=75pt]{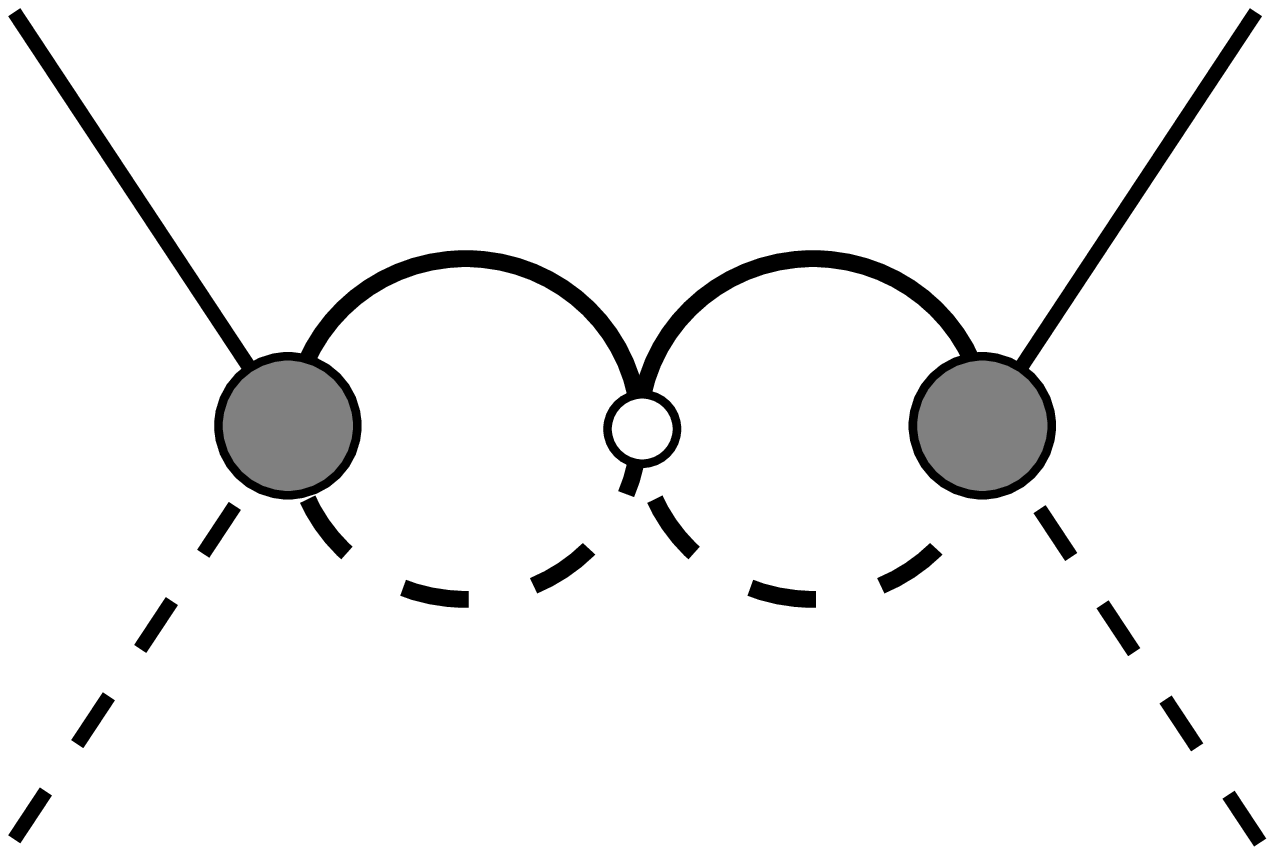}}
\end{center}

%caption: bilocal operator matrix elements
\caption{Diagrams for the matrix element of the operator $\psi_{\uparrow}^{\dagger} \psi_{\downarrow}^{\dagger} \psi_{\downarrow} \psi_{\uparrow}$}
\label{fig:contact-matrixelements}
\end{figure}
%end: figure contact diagrams
%%%%%%%%%%%%%%%%%%%%%%%%%%%%%
To determine the Wilson coefficient of the operator $\psi_{\uparrow}^{\dagger} \psi_{\downarrow}^{\dagger} \psi_{\downarrow} \psi_{\uparrow}$, we compute its matrix element between the two-particle scattering states, which is given by the sum of the four diagrams shown in figure \ref{fig:contact-matrixelements}. Using the Lippmann-Schwinger equation in the form
\begin{equation}
 \label{eq:LSE}
 - \frac{\mathcal{A}(E)}{g_1} = 1 + i \mathcal{A}(E)\frac{m_r}{|p|},
\end{equation}
where the factor $m_r/|p|$ is the value of the loop integral, the sum of the four diagrams can be reexpressed and the matrix element is
\begin{equation}
 \label{eq:contact-matrixelement}
 \bra{\pm p'} \psi_{\uparrow}^{\dagger} \psi_{\downarrow}^{\dagger} \psi_{\downarrow} \psi_{\uparrow}(R) \ket{\pm p} = \frac{\mathcal{A}(E) \mathcal{A}(E')}{g_1^2}.
\end{equation}
This matches the $|x|^3$-term in the expansion of \eqref{eq:2p-matching-lhs-nonanalytic} with Wilson coefficient $g_1^2 m_r^2 |x|^3 / 3$. In summary, the short distance OPE \eqref{eq:OPE-1pdm-general} of the one-particle density matrix, up to order $x^4$, is the sum of three analytic contributions and a leading order non-analytic contribution proportional to $|x|^3$. The operator responsible for this non-analyticity is 
$\psi_{\uparrow}^{\dagger} \psi_{\downarrow}^{\dagger} \psi_{\downarrow} \psi_{\uparrow}(R)$, in complete analogy to the 
three-dimensional case~\cite{Braaten_2008_short}. Explicitly, the Wilson-OPE for the one-particle density matrix up to order $x^3$ reads
\begin{equation}
 \label{eq:OPE-1pdm-explicit}
	\psi^{\dagger}_{\sigma} (R) \psi_{\sigma} (R+x) = 
	\sum_{m=0}^3 \frac{x^m}{m!} \cdot \psi_{\sigma}^{\dagger} \left( \partial_R^m \psi_{\sigma} \right)(R) + \frac{1}{3} m_r^2 g_1^2 |x|^3 \cdot \psi_{\uparrow}^{\dagger} \psi_{\downarrow}^{\dagger} \psi_{\downarrow} \psi_{\uparrow} (R) + \mathcal{O}(x^4).
\end{equation}
Clearly, the zeroth order term at $x=0$ is just the density of species $\sigma$ at location $R$, as expected. The second order Wilson coefficient is associated with the operator $\mathcal{O}_{\sigma,2}(R)=\psi_{\sigma}^{\dagger} \left( \partial_R^2 \psi_{\sigma} \right)(R)$
which is the kinetic energy density. Expanding the one-particle density matrix in the ground state of a uniform, balanced Fermi gas as $\langle \psi^{\dagger}_{\sigma}(x) \psi_{\sigma} (0) \rangle = n(1+c_2 n^2 x^2+...)/2$, this contribution can be expressed in terms of the dimensionless ground state energy $e(\gamma)$ introduced in \eqref{eq:groundstate-energy} below in the form
\begin{equation}
 \label{eq:c-two}
c_{2} = -[e(\gamma) - \gamma e'(\gamma)]/2 \, , 
\end{equation}
 %when we expand g1 as 1+nc_1|x|+n^2c_2|x|^2+.. n is the total density
in close analogy to a result derived by Olshanii and Dunjko~\cite{Olshanii_2003_1D} for the Lieb-Liniger gas. Since this is an analytic contribution in
$x$, it does not give rise to a power law in the associated momentum distribution.

The leading non-analytic contribution appears only at third order in $x$ unlike in three dimensions, where it is of order $r=|\mathbf{x}|$. The absence in 1D of a  contribution proportional to $|x|$  may be traced back to the fact that both the kinetic and the interaction energy are finite in 1D. A term $\sim |x|$ in the
one-particle density matrix would lead to a $1/k^2$-tail in the momentum distribution, thus making the kinetic energy divergent. The leading term in the momentum distribution for high $k$ can be obtained by inserting the OPE \eqref{eq:OPE-1pdm-explicit} up to order $|x|^3$ into equation \eqref{eq:momentumdistribution-OPE}.  The resulting power law tail is obtained from the formal Fourier transform 
\begin{equation}
 \label{eq:FT-absr-alpha}
 \int dx |x|^{\alpha} e^{-ikx} = 2 \Gamma ( \alpha + 1) \cos \left[  \frac{\pi}{2}(\alpha+1) \right]  \frac{1}{|k|^{\alpha +1}},
\end{equation}
of $|x|^{\alpha}$, with $\alpha$ a non-even real number. Specifically, for $\alpha = 3$ one obtains $12/k^4$. With the prefactors given by 
\eqref{eq:OPE-1pdm-explicit}, we have thus shown that the momentum distribution
\begin{equation}
 \label{eq:asymptotic-momentum-distribution}
 \tilde{n}_{\sigma}(k) \stackrel{k \rightarrow \infty}{\longrightarrow} \frac{C}{k^4}
\end{equation}
of a two-component Fermi gas in one dimension with contact interactions,
exhibits a power law decay $C/k^4$ at large momenta $k$. (In practice, for a 
uniform gas with 1D density $n$, the asymptotic behavior shows up at $k\gg n$). 
The constant prefactor $C$, called the contact, is the same for both species and is positive by
definition. Microscopically it is given by the expression 
\eqref{eq:contact-exact} which is proportional to the expectation value $\langle H'\rangle$ of the interaction 
energy. In the translational invariant case,  $\langle H'\rangle\sim L$ is extensive since the 
pair distribution function \eqref{eq:gupdown} only depends on the coordinate difference. 
The resulting intensive contact density $ \mathcal{C}=C/L$ is then given by
\begin{equation}
 \label{eq:contact-translational-invariant}
 \mathcal{C} = \frac{4}{a_1^2} n_{\uparrow} n_{\downarrow} g_{\uparrow\downarrow}^{(2)}(0).
\end{equation}
This result is completely analogous to the expression
\begin{equation}
\label{eq:bosonic-contact}
  \mathcal{C_B} = \frac{4}{a_1^2} n^2 g^{(2)}(0)
\end{equation}
for the contact density of the ground state of the uniform Lieb-Liniger gas derived by Olshanii and Dunjko~\cite{Olshanii_2003_1D}.  Note that in the non-interacting case $g_1=0$, the non-analytic terms disappear and there is no tail, since the momentum distribution is just a Fermi-Dirac or
Bose-Einstein distribution for free particles which exhibits no $1/k^4$ tail at any temperature.

%%%%%%%%%%%%%%%%%%%%%%%%%%%%%%%%%%%%%%%%%%%%%%%%%%%%%%%%%%%%%
%%Derivation of the Tan energy and pressure relation
\section{Energy and Pressure Relation} \label{sec:energy}
In the following we will show that the connection between the asymptotic behavior of the momentum distribution and the interaction part
of the Hamiltonian also allows to derive three other universal relations for the two-component Fermi gas with contact interactions 
which are again analogous to those in the 3D  case.  We start with the energy relation:

The total energy $E = \langle H \rangle = \langle T \rangle + \langle I \rangle + \langle \mathcal{V} \rangle$  of the Fermi gas in the presence of an additional 
one-body confining potential $\mathcal{V}(R)$ can be expressed in terms of the contact $C$ and the momentum distribution $\tilde{n}_{\sigma}(k)$, as
\begin{equation}
 \label{eq:energy-relation}
 E = \sum_{\sigma} \int \frac{dk}{2 \pi} \frac{k^2}{2m_{\sigma}} \tilde{n}_{\sigma} (k) - a_1 \frac{C}{4 m_r} +\langle \mathcal{V} \rangle.
\end{equation}
This relation follows quite simply by inserting the definition \eqref{eq:contact-exact} of the contact into the Hamiltonian 
\eqref{eq:hamilton-density}, and rewriting the kinetic energy in momentum space
\begin{equation}
 \label{eq:kinetic-terms}
 \int dR \Braket{\partial_R \psi^{\dagger}_{\sigma}(R) \partial_R \psi_{\sigma}(R)}
 =
 \int \frac{dk}{2\pi} k^2 \underbrace{\Braket{\tilde{\psi}_{\sigma}^{\dagger}(k) \tilde{\psi}_{\sigma}(k)}}_{\tilde{n}_{\sigma}(k)}.
\end{equation}
The energy relation expresses the energy as a functional of the momentum and -- in the inhomogeneous case -- also the density distribution. 
It serves as a basis from which, together with the adiabatic relation, other Tan relations will follow. A special case of the energy relation 
is the virial theorem:

For an harmonic trapping potential $\mathcal{V}(R)= m \omega^2 R^2 /2$, the energy can be written as the sum of the trapping energy 
and the contact in the following way:
\begin{equation}
\label{eq:virial-theorem}
 E = 2 \langle\mathcal{V} \rangle - a_1 \frac{C}{8m_r}.
\end{equation}
This relation can be derived by a simple scaling argument. The only three energy scales are $\omega$, $1/m_ra_1^2$ and $k_B T$.
%{\it Caution: This argument is correct only at zero temperature. Thus our derivation of the virial theorem only 
%applies in the ground state. In fact, however, the theorem should also be true at arbitrary finite temperature!}
The free energy
\begin{equation}
\label{eq:freeenergy-dimlessfct}
 F(T,N_{\uparrow},N_{\downarrow},\omega, a_1) = \omega f \left(\frac{k_B T}{\omega}, \frac{\omega}{1/ma_1^2} ,N_{\uparrow},N_{\downarrow} \right)
\end{equation}
can therefore be expressed in terms of a dimensionless function $f$, which depends on the ratios $k_BT / \omega$, $\omega / (1/ma^2)$ and the two particle numbers $N_{\uparrow},N_{\downarrow}$. From \eqref{eq:freeenergy-dimlessfct}, one can deduce the simple scaling law
\begin{equation}
 \label{eq:freeenergy-scaling-viral}
 F(\lambda T, N_{\uparrow},N_{\downarrow},\lambda \omega, \lambda^{-\frac{1}{2}}a_1) = \lambda F(T,N_{\uparrow},N_{\downarrow},\omega, a_1).
\end{equation}
The derivative of equation \eqref{eq:freeenergy-scaling-viral} with respect to $\lambda$ at $\lambda = 1$ yields
\begin{equation}
 \label{eq:freeenergy-de-viral}
\left( T \frac{\partial}{\partial T} + \omega \frac{\partial}{\partial \omega} - \frac{1}{2}a_1 \frac{\partial}{\partial a_1} \right) F = F,
\end{equation}
where all the partial derivatives are to be understood as leaving all other system variables constant. Since the free energy is just the Legendre transform of the energy, its partial derivatives at constant temperature $T$ with respect to $\omega$ and $a_1$ are equal to those of the energy at the 
associated value of the entropy. Therefore, using $\partial F / \partial T = -S$, the energy turns out to obey the differential equation
\begin{equation}
\label{eq:energy-diffop}
 \left( \omega \frac{\partial}{\partial \omega} - \frac{1}{2} a_1 \frac{\partial}{\partial a_1} \right)E=E.
\end{equation}
 This leads immediately to the relation \eqref{eq:virial-theorem} by using the Hellmann--Feynman theorem with the trapping frequency $\omega$ as a parameter of the Hamiltonian and the adiabatic relation \eqref{eq:adiabatic-relation}.  Again, the virial theorem is analogous to a corresponding 
relation in 3D that was derived first by Thomas \textit{et al.} ~\cite{Thomas_2005_virial}  for the unitary gas at $a=\infty$ and was 
extended to finite scattering lengths by Tan ~\cite{Tan_2008_virialpressure}. Note that at unitarity or -- correspondingly --  at $a_1=0$ in
1D, the virial theorem allows to extract the total energy from the density profile $n(R)$ in the harmonic trap.

Finally, for the homogeneous (i.e. $\mathcal{V}_{\sigma} (R) = 0$) two-component Fermi gas with
contact interactions, there is a pressure relation which connects
 pressure and energy density  $\mathcal{E}$ (i.e. energy per length) by
\begin{equation}
\label{eq:pressure-relation}
 p = 2 \mathcal{E} + a_1 \frac{\mathcal{C}(a_1)}{4m_r}\, ,
\end{equation}
where $\mathcal{C} = C / L$ is the contact density.
For its derivation, we consider the free energy density $\mathcal{F}$ of the homogenous system. Since $\epsilon_{F\uparrow}\sim n_{\uparrow}^2$ 
and $n_{\uparrow} \epsilon_{F\uparrow}$ are characteristic scales for energy and energy density,  dimensional analysis requires 
that  $\mathcal{F}$ can be expressed in terms of a dimensionless function
\begin{equation}
\label{eq:free-energy-dimlessfct}
 \mathcal{F}(T,a_1,n_{\uparrow},n_{\downarrow}) = \alpha n_{\uparrow}^3 f \left( \alpha_T T/ n_{\uparrow}^2, n_{\uparrow} a_1, 
 n_{\downarrow}/ n_{\uparrow}  \right)\, ,
\end{equation}
where $n_{\sigma}$ is the density of species and $n_{\uparrow}>0$ without restriction of generality.
% and $T_{F,\sigma}$ and $\epsilon_{F,\sigma}$ are the corresponding Fermi temperatures and Fermi energies. 
The function $f$ depends on three dimensionless arguments which set the scale for temperature, interaction strength and 
a possible imbalance of the densities.  The constants $\alpha$ and $\alpha_T$ carry  factors that are needed
to make the arguments dimensionless and are irrelevant for the following.  Apparently, equation \eqref{eq:free-energy-dimlessfct} 
implies the following scaling-behavior of the free energy density
\begin{equation}
 \label{eq:free-energy-scaling}
 \mathcal{F}(\lambda^2 T, \lambda^{-1} a_1, \lambda n_{\uparrow}, \lambda n_{\downarrow}) = \lambda^3 \mathcal{F}(T,a_1,n_{\uparrow},n_{\downarrow}),
\end{equation}
for an arbitrary, dimensionless parameter $\lambda$. Taking the derivative of equation \eqref{eq:free-energy-scaling} with respect to the parameter $\lambda$ at  $\lambda = 1$ yields
\begin{equation}
\label{eq:free-energy-diffop}
\left[ 2T\frac{\partial}{\partial T} +n_{\uparrow} \frac{\partial}{\partial n_{\uparrow}} +n_{\downarrow}\frac{\partial}{\partial n_{\downarrow}} - a_1 \frac{\partial}{\partial a_1} \right] \mathcal{F} =  3 \mathcal{F}.
\end{equation}
Now, quite generally, the grand canonical potential per length for a two-component system is given by
\begin{equation}
 \mathcal{J} = -p = \mathcal{F} - n_{\uparrow} \mu_{\uparrow} - n_{\downarrow} \mu_{\downarrow}\, .
\end{equation}
This allows to replace the free energy density $\mathcal{F}$ in equation \eqref{eq:free-energy-diffop} by $n_{\uparrow} \mu_{\uparrow} + n_{\downarrow} \mu_{\downarrow} -p$. Using the thermodynamic relations $\partial_T \mathcal{F} = -s$, where $s$ is the entropy density, and $\partial_{n_{\sigma}} \mathcal{F} = \mu_{\sigma}$ results, after rearranging the terms, in
\begin{equation}
\label{eq:free-energy-uptopartial}
 p = 2 \underbrace{\left( n_{\uparrow} \mu_{\uparrow} + n_{\downarrow} \mu_{\downarrow} -p + T s \right)}_{ = \mathcal{F} + Ts = \mathcal{E}} + a_1 \frac{\partial}{\partial a_1} \mathcal{F}.
\end{equation}
Since the partial derivatives in equations \eqref{eq:free-energy-diffop} and \eqref{eq:free-energy-uptopartial} are to be understood as leaving the other system variables constant, we can replace the $\partial_{a_1} \mathcal{F}$ term by the adiabatic relation \eqref{eq:adiabatic-relation} for the energy density. The resulting equation is the pressure relation \eqref{eq:pressure-relation}. Note that the sign of the correction term is opposite to that in the three-dimensional pressure relation ~\cite{Tan_2008_virialpressure}. At first sight, it therefore seems that a repulsive interaction $a_1 < 0$ lowers the 
pressure and vice versa. However, upon expressing $2 \mathcal{E}$ in terms of a kinetic and an interaction energy part using the energy relation 
\eqref{eq:energy-relation}, we see that the interaction energy per length is equal to $-a_1 \mathcal{C} / 4m_r$, which is positive for repulsive interactions, 
and negative for attractive ones, as expected.

A quite nontrivial consequence of the pressure relation arises by considering the limit $a_1=0$ i.e. infinite interaction strength.
In this limit equation \eqref{eq:pressure-relation} predicts that pressure and energy density are related in precisely the same manner 
as  for an {\it ideal} gas of non-relativistic particles in one dimension (note that this requires the product 
$a_1\mathcal{C}(a_1)\sim g_{\uparrow\downarrow}^{(2)}(0)/a_1$ to vanish in the limit $a_1\to 0^{-}$, 
which is indeed the case as will be discussed in section 7 below). This surprising conclusion is a consequence of the fact 
that a 1D Fermi gas with contact interactions is scale invariant at infinite interaction strength $a_1=0$. 
The underlying reason why an ideal gas type equation of state appears in this limit can be understood 
in simple terms by noting that the balanced  two-component Fermi gas at infinite repulsion behaves like a free Fermi 
gas with a doubled value of the Fermi wave vector. Effectively, an infinite zero range repulsion in the two-component 
system is equivalent to a Pauli exclusion principle also between fermions of opposite spin. The resulting
equation of state is thus $p=2\mathcal{E}$ at arbitrary temperatures despite the fact that it is
a strongly interacting system. The same holds true in the case of infinite attraction $a_1\to 0^+$
because the two-component Fermi gas in this limit is a Tonks-Girardeau gas of bosonic dimers, i.e. effectively
an ideal single-component Fermi gas with $\tilde{k}_F=\pi n/2$~\cite{Fuchs_2004, Tokatly_2004}. 
Note that in this case, both the energy density $\mathcal{E}$ 
and the contact density $\mathcal{C}$ contain an infinite constant arising from the two-body binding energy, which diverges as $g_1\to -\infty$.  
This constant, however, cancels in the equation of state, which -- at infinite attraction -- is again the same as if there were no interactions at all.\\

%%%%%%%%%%%%%%%%%%%%%%%%%%%%%%%%%%%%%%%%%%%%%%%%%%
%Density-Density correlator including OPE matching
\section{Short distance expansion of the pair distribution function} \label{sec:dd-correlator}
In our discussion so far, we have only used the behavior of the one-particle density matrix at short distances. Its leading 
non-analytic contribution defines the contact and gives rise to Tan relations that are closely related to those in the 3D case 
~\cite{Tan_2008_energetics, Tan_2008_adiabatic, Braaten_2008_short} and also to those for 1D Bose gases~\cite{Olshanii_2003_1D}. 
In the following we will show that the contact also arises as a non-analytic contribution in the OPE of the {\it two-particle} density matrix 
which gives additional relations for the pair distribution function and the related static structure factor.  Specifically we will derive the 
following short distance expansion 
\begin{equation}
 \label{eq:dd-correlator-OPE}
 \begin{split}
  \hat{n}_{\uparrow} \left( R-\frac{x}{2} \right) \hat{n}_{\downarrow}  \left(R + \frac{x}{2}\right) = &
 \left( 1 - \frac{2|x|}{a_1} \right) \psi^{\dagger}_{\uparrow}\psi^{\dagger}_{\downarrow}\psi_{\downarrow}\psi_{\uparrow}(R) \\ &  + \frac{x}{2} \cdot \left[ \hat{n}_{\uparrow} \partial_R \hat{n}_{\downarrow} - ( \partial_R \hat{n}_{\uparrow} ) \hat{n}_{\downarrow} \right] (R) + \mathcal{O}(x^2)
\end{split}
\end{equation}
of the $\uparrow - \downarrow$ density--density correlator up to linear order in the separation $x$. 
%%%%%%%%%%%%%%%%%%%%%%%%%%%%%%%%%%%%%%
%begin: figure d-d correlator diagrams
\begin{figure}

\begin{center}
\subfigure[{ }]{\label{subfig:dd1} \includegraphics[width=80pt,height=75pt]{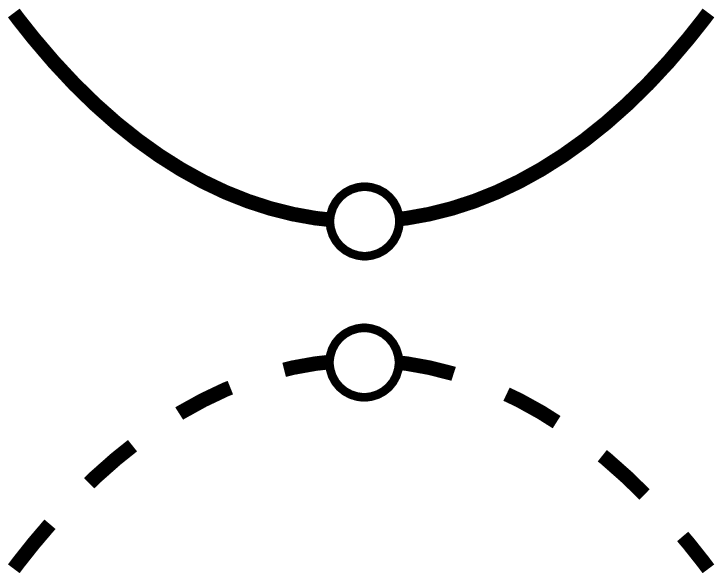}} \hspace{5pt}
\subfigure[{ }]{\label{subfig:dd2} \includegraphics[width=50pt,height=75pt]{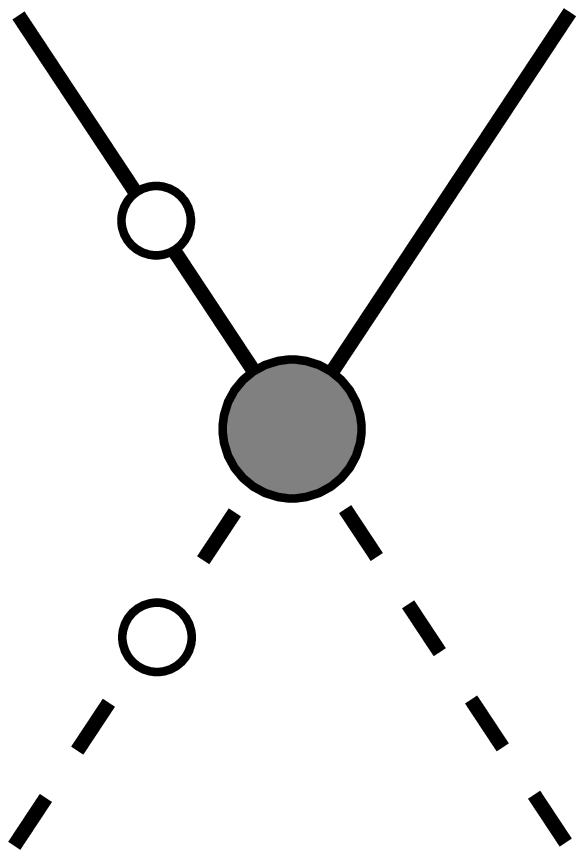}} \hspace{5pt}
\subfigure[{ }]{\label{subfig:dd3} \includegraphics[width=50pt,height=75pt]{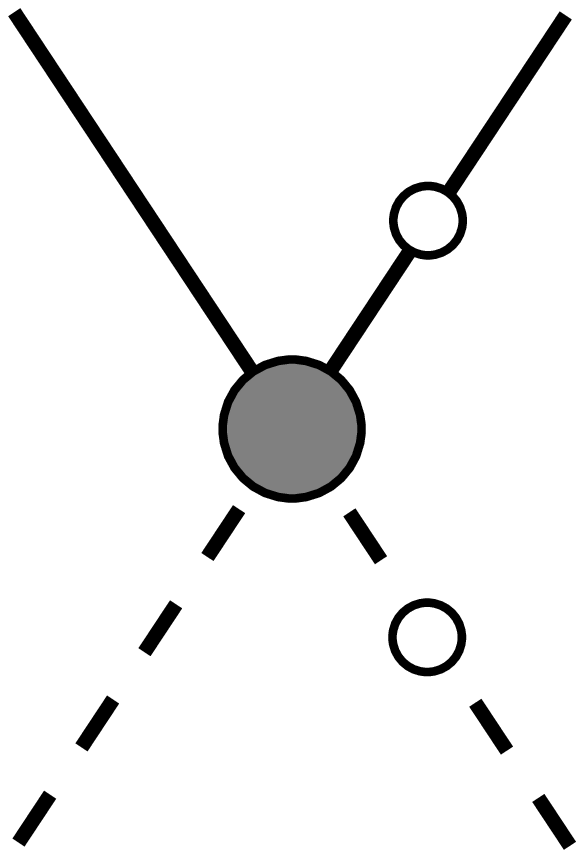}} \hspace{5pt}
\subfigure[{ }]{\label{subfig:dd4} \includegraphics[width=111pt,height=75pt]{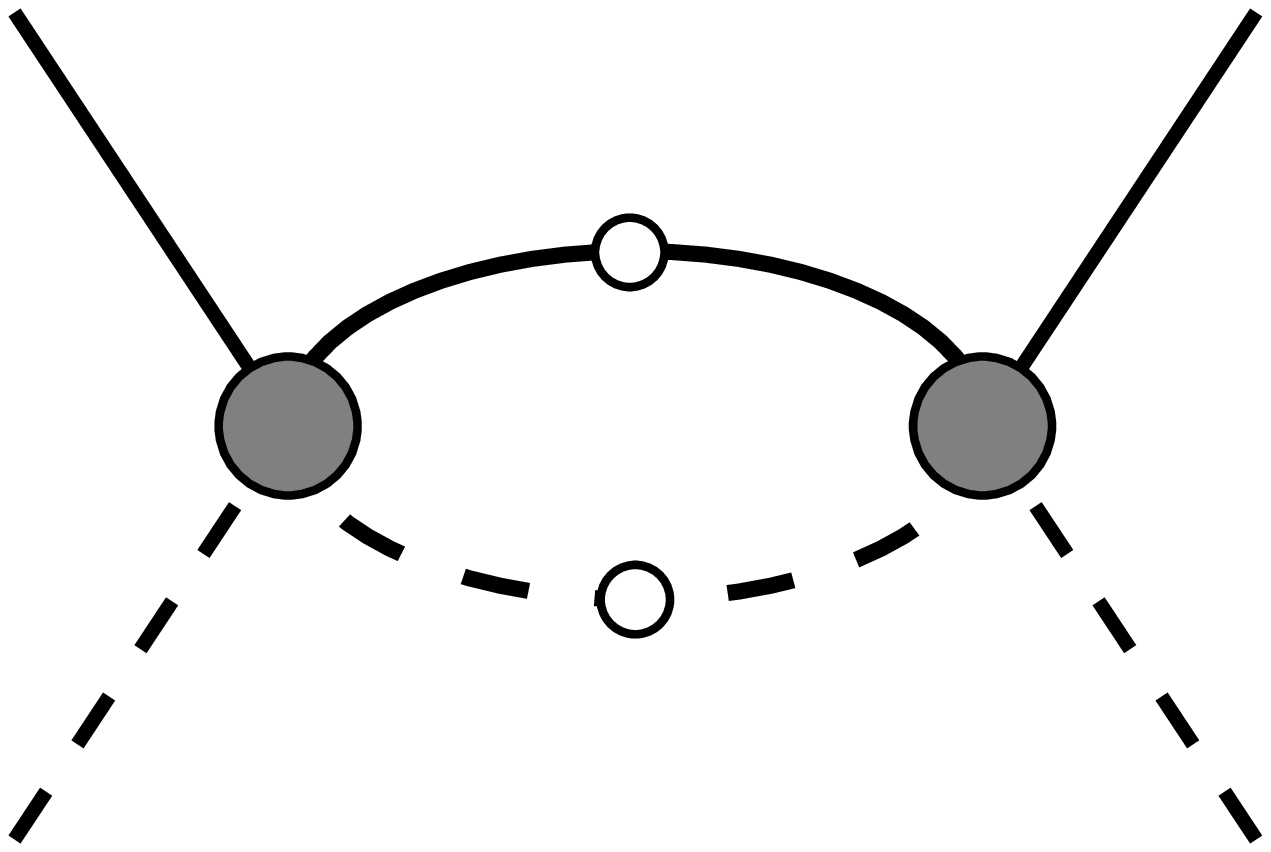}}
\end{center}

%caption: bilocal operator matrix elements
\caption{Diagrams for the matching of the density--density correlator on the left hand side of its OPE.}
\label{fig:dd-matrixelements}
\end{figure}
%end: figure d-d correlator diagrams
%%%%%%%%%%%%%%%%%%%%%%%%%%%%%%%%%%%%
The fact that the $x \rightarrow 0$ limit of the pair correlator is simply the contact operator $\mathcal{C}(R)$ is not surprising. Note, however, that in 
three dimensions the relation is more complicated and reads~\cite{Tan_2008_energetics, Braaten_2008_short}
\begin{equation}
 \label{eq:dd-correlator-3D}
\langle  \hat{n}_{\uparrow} \left(\mathbf{R}-\frac{\mathbf{x}}{2} \right) \hat{n}_{\downarrow}  \left(\mathbf{R} + \frac{\mathbf{x}}{2}\right)\rangle =
\frac{\mathcal{C}(\mathbf{R})}{16\pi^2\vert\mathbf{x}\vert^2}+\ldots
\end{equation}
due to anomalous scaling of the contact operator. Equation \eqref{eq:dd-correlator-OPE} shows that in 1D, the contact is also responsible for the 
leading order {\it non-analytic} contribution to the pair distribution function, which is proportional to $|x|$. 

To prove the OPE given in equation \eqref{eq:dd-correlator-OPE}, first note that one cannot simplify the matching of some operators in one-particle states, as was the case for the OPE of the one-particle density matrix. Thus, we use scattering states $\bra{\pm p'}$ and $\ket{\pm p}$ to compute the matrix elements of the operators on both sides. The left hand side is the sum of the four diagrams shown in figure \ref{fig:dd-matrixelements}. Evaluating these yields
\begin{equation}
\label{eq:dd-matching-lhs}
  e^{-i(p-p')x} +  m_r i \mathcal{A}(E')e^{-ipx} \frac{e^{i|p'||x|}}{|p'|} + m_r i \mathcal{A}(E)e^{ip'x} \frac{e^{i|p||x|}}{|p|} + m_r^2 i \mathcal{A}(E) i \mathcal{A}(E') \frac{e^{i|p||x|}}{|p|} \frac{e^{i|p'||x|}}{|p'|}\, .
\end{equation}
Expanding this expression to linear order $x$, there are analytic contributions proportional to $x^0$ and $x^1$ plus a non-analytic one of order $|x|$. The $x^0$-term can be rewritten, using the Lippmann-Schwinger equation \eqref{eq:LSE} as $\mathcal{A}(E) \mathcal{A}(E') / g_1^2$. The non-analytic $|x|$-term is given by
\begin{equation}
 \label{eq:dd-matching-lhs-naterm}
 \left[ i \mathcal{A}(E) \left( 1 + i \mathcal{A}(E') \frac{m_r}{|p'|} \right) + i \mathcal{A}(E') \left( 1 + i \mathcal{A}(E) \frac{m_r}{|p|} \right) \right] \cdot i m_r |x| = \frac{2 m_r |x|}{g_1} \mathcal{A}(E) \mathcal{A}(E'),
\end{equation}
where we have again used the Lippmann-Schwinger equation \eqref{eq:LSE}. The sum of the $x^0$ and the $|x|$ parts now reads
\begin{equation}
 \label{eq:dd-matching-contact}
 (1+2 m_r g_1 |x|)\frac{\mathcal{A}(E) \mathcal{A}(E')}{g_1^2}.
\end{equation}
Recall, that $\mathcal{A}(E) \mathcal{A}(E') / g_1^2$ is just the matrix element of the contact operator $\psi^{\dagger}_{\uparrow}\psi^{\dagger}_{\downarrow}\psi_{\downarrow}\psi_{\uparrow}(R)$ in these states, which is given in equation \eqref{eq:contact-matrixelement}. The Wilson coefficient of the contact operator, up to order $|x|$, is therefore given by $ (1 + 2 m_r g_1 |x|)$. We abbreviate the matching of the analytic part by noting that the matrix element of the operator $\hat{n}_{\uparrow} \partial_R \hat{n}_{\downarrow} - ( \partial_R \hat{n}_{\uparrow} ) \hat{n}_{\downarrow}$ between the scattering states, which can again be represented by the diagrams in figure \ref{fig:contact-matrixelements}, is given by
\begin{equation}
\label{eq:dd-matching-analytic}
 2 \left[ -i(p-p') + (-ip) i\mathcal{A}(E') \frac{m_r}{|p'|} + (ip') i\mathcal{A}(E) \frac{m_r}{|p|} \right]\, .
\end{equation}
This matches exactly the linear term in the expansion of equation \eqref{eq:dd-matching-lhs} with Wilson coefficient $x/2$, thus completing the
derivation of the OPE \eqref{eq:dd-correlator-OPE} of the density--density correlator.

In order to discuss the consequences of the short distance singularity in equation \eqref{eq:dd-correlator-OPE} for physical
observables, we consider the full pair distribution function $g^{(2)}(R,R')$. Quite generally, it is defined by the density--density correlation via 
\begin{equation}
 \langle\hat{n}(R)\hat{n}(R')\rangle= n(R)n(R') g^{(2)}(R,R')+\delta(R - R')n(R)\, .
 \label{eq:g_2}
\end{equation} 
Considering, for simplicity, the translation invariant case where
the pair distribution function only depends on the separation $x=R'-R$ of the two particles,  the total pair distribution 
function 
\begin{equation}
    g^{(2)}(x)=\frac{1}{n^2}\left( n_{\uparrow}^2\, g_{\uparrow\uparrow}^{(2)}(x)+n_{\downarrow}^2\, g_{\downarrow\downarrow}^{(2)}(x)+
    2n_{\uparrow}n_{\downarrow}\, g_{\uparrow\downarrow}^{(2)}(x)\,\right)
 \label{eq:g_2sigma}
\end{equation} 
of a two-component Fermi gas splits into separate spin contributions $g_{\sigma,\sigma'}^{(2)}(x)$ that all approach unity 
as $x\to\infty$. For small separations $x\to 0$, the pair distribution function for equal spins will vanish like $x^2$ because
the antisymmetry due to the Pauli principle forces the many-body wave-function to vanish linearly $\sim (x_i-x_j)$ as two coordinates $x_i$ and $x_j$ of fermions with equal spin approach each other. As a result, equation \eqref{eq:dd-correlator-OPE} implies
that the total pair distribution function at short distances is given by
\begin{equation}
    g^{(2)}(x)= \frac{2n_{\uparrow}n_{\downarrow}}{n^2}\, g_{\uparrow\downarrow}^{(2)}(0) \left(1- \frac{2|x|}{a_1} +\mathcal{O}(x^2) \right)
 \label{eq:g_2short}
\end{equation}
up to linear order in $x$. For repulsive interactions $a_1<0$, therefore, the pair distribution function rises
linearly from its value at zero separation. The associated short distance singularity $\sim |x|$ gives rise to a $1/q^2$ power law 
tail in the associated static structure factor
\begin{equation}
 S(q) = 1 + n \int dx \, e^{-iqx} \left[ g^{(2)} (x) - 1 \right]\, .
 \label{eq:structurefactor-def}
\end{equation}
Introducing a dimensionless coupling strength $\gamma=-2/(na_1)$, the Fourier transform of the non-analytic contribution in 
\eqref{eq:g_2short} leads to the asymptotic behavior
in the form
\begin{equation}
 S(q\to\infty)=1-\gamma g_{\uparrow\downarrow}^{(2)}(0)\cdot\frac{4n_{\uparrow}n_{\downarrow}}{q^2}+\ldots\, .
  \label{eq:structurefactor-asymptotic}
\end{equation}
The tail of the static structure factor is therefore proportional to the contact density in equation  \eqref{eq:contact-translational-invariant}.
A similar connection holds in 3D as pointed out by Hu et. al.~\cite{Hu_2010_structurefactor}. In the 3D case, however,
the tail is proportional to $\mathcal{C}/q$, which is  a result of the $1/r^2$-singularity at short distances described by equation 
\eqref{eq:dd-correlator-3D}. Note that the tail in  \eqref{eq:structurefactor-asymptotic} vanishes in the infinite repulsion limit $\gamma\to\infty$ 
because $ g_{\uparrow\downarrow}^{(2)}(0)\sim 1/\gamma^2$, as will be discussed in section 7 below. At $\gamma=\infty$,
the pair distribution function will in fact be equal to that of a non-interacting single-component Fermi gas with Fermi wave vector
$\tilde{k}_F=2k_F=\pi n$, while the momentum distribution $n_{\sigma}(k)$ will still have the nontrivial form of a Luttinger liquid
(see section 7 below).\\
 
 Finally, we note that in 1D the contact density also arises in a sum rule for the static structure factor. Indeed, the Fourier inversion  
\begin{equation}
 g^{(2)} (x) = 1 + \frac{1}{n} \int \frac{dq}{2\pi} e^{iqx} \left[ S(q) - 1 \right]
 \label{eq:g_2S(q)}
\end{equation}
of equation \eqref{eq:structurefactor-def} at $x=0$ immediately implies a sum rule
\begin{equation}
 \frac{2n_{\uparrow}n_{\downarrow}}{n^2}\, g_{\uparrow\downarrow}^{(2)}(0) = 1 + \frac{1}{n} \int_0^{\infty} \frac{dq}{\pi} \left[ S(q) - 1 \right]
 \label{eq:structurefactor-sumrule}
\end{equation}
which connects the integrated static structure factor with the contact density. Here, we have used the fact that the contributions to the pair 
distribution function for equal spins vanish at $x=0$ and that $S(q)$ is an even function. In contrast to 3D, where $g^{(2)}(0)$ is 
not defined because even the zeroth moment of $S(q)-1$ diverges due to the $1/q$-tail, the finiteness of the pair distribution
function at vanishing distance in 1D guarantees that the deviation of the static structure factor from its trivial limit $S^{(0)}(q)=1$ 
that appears for both an ideal classical gas or an ideal Bose-Einstein condensate, is integrable.

%%%%%%%%%%%%%%%%%%%%%%%%%%%%%%%%%%%%%%%%%%%%%%%%%%%%%%%%%%%%%
%%Contact density for uniform gases and the infinite repulsion limit
\section{Contact density for uniform gases} \label{sec:contact}
As mentioned in the introduction, the Tan relations that have been derived above
are valid for an arbitrary state, both at zero and at finite temperature and in the presence of a
non-vanishing one-body potential $\mathcal{V}(R)$. The simplest case of interest, for which explicit results for 
the contact density at arbitrary coupling strengths can be given on the basis of the Bethe Ansatz solution,  
is the ground state of a uniform gas with total density $n$, in which the two spin components are equally populated. 
In the repulsive case, this gas is known to be a Luttinger liquid~\cite{Haldane_1981, Giamarchi}.  Quite generally,
the one-particle density matrix of a Luttinger liquid of fermions decays with
a power law $\sim x^{-(\alpha+1)}\exp{ik_Fx}$ at zero temperature. The 
resulting momentum distribution therefore exhibits a singularity $\sim\vert k-k_F\vert^{\alpha}$ at the 
bare Fermi wave vector $k_F$, replacing the simple jump from one 
to zero of an ideal Fermi gas. The exponent $\alpha=(K-1)^2/2K$ is related the Luttinger parameter $K$ 
which -- for the case of contact interactions -- continuously varies between $K=1$ for the non-interacting gas 
and $K=1/2$ for infinite repulsion~\cite{Giamarchi}.   
For attractive interactions, the balanced two-component Fermi gas is a Luther-Emery liquid~\cite{Luther_1974, Fuchs_2004}.
It has a finite gap for spin excitations due to the appearance of bound pairs with opposite spin, 
thus eliminating the singularity of the momentum distribution at $k_F$. 
In both cases, the strength of the correlations in the ground state is characterized by a dimensionless coupling 
constant $\gamma\equiv -2/na_1$, which is inversely proportional to the  density. In 1D, the strong coupling limit $|\gamma|\gg 1$  
is therefore reached at {\it low} densities. This somewhat counterintuitive fact  can be understood by noting that low 
densities imply small momenta and, moreover, the 1D scattering amplitude (\ref{eq:reflection}) has its maximum phase 
shift $\delta(0)=\pi/2$ as $k\to 0$ because 1D potentials become impenetrable at 
zero energy, i.e. $r(k\to 0)=-1$. \\

The ground state energy of the uniform gas can be determined from the solution of the
Gaudin-Yang integral equations, which can in fact be solved for arbitrary polarization
$(n_{\uparrow}-n_{\downarrow})\ne 0$~\cite{Batchelor_2006}. In the case of a balanced gas
with equal masses, the ground state energy  per particle
\begin{equation}
 \label{eq:groundstate-energy}
\frac{E_0}{N}=\frac{\hbar^2 n^2}{2m}\cdot e(\gamma)
\end{equation} 
is conveniently expressed in terms of a dimensionless function $e(\gamma)$. For a repulsive gas with
$\gamma\geq 0$, the function $e(\gamma)$ increases monotonically from the value $e(0)=\pi^2/12$ that 
corresponds to the non-interacting two-component Fermi gas in 1D to $e(\infty)=\pi^2/3$. In the attractive case
$\gamma<0$, it is convenient to subtract a contribution $-\gamma^2/4$ from the two-body binding energy.
The relevant many-body energy $\tilde{e}(\gamma\leq 0)=e(\gamma\leq 0)+\gamma^2/4$ then decreases monotonically
from $\tilde{e}(0)=\pi^2/12$ to $\tilde{e}(-\infty)=\pi^2/48$~\cite{Krivnov_1975, Fuchs_2004, Tokatly_2004}.   
In the case of a balanced gas, the connection \eqref{eq:Hellman-Feynman} between the derivative of the energy per
length with respect to the coupling constant $g_1$ and the pair distribution function at vanishing separation 
reduces to the simple relation
\begin{equation}
 \label{eq:eprime}
g_{\uparrow \downarrow}^{(2)}(0)=2e'(\gamma)\, ,
\end{equation} 
where $e'(\gamma)$ denotes the derivative of the dimensionless ground state energy defined in equation (\ref{eq:groundstate-energy}).

%%%%%%%%%%%%%%%%%%%%%%%%%%%%%%%%%%%%%%%%%%%%%%%%%%%%%%%%%%
%%%Begin: Figure of the g2 plot and s plot
\begin{figure}
 \begin{center}
 \includegraphics[width=0.8\columnwidth]{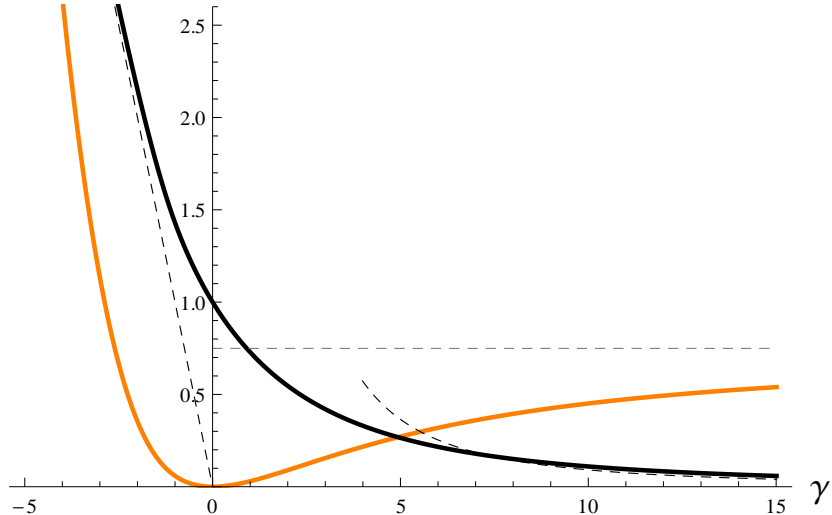}
 \end{center}
\caption{The pair distribution function $g_{\uparrow\downarrow}^{(2)}(0)$ at zero distance (black curve) and the associated dimensionless contact density 
$s$ (orange curve) in the ground state of a balanced homogeneous Fermi gas as a function of the dimensionless coupling $\gamma = -2/na_1$. The black dashed line $-\gamma$ on the left originates from the binding energy of two atoms, the right dashed curve follows from the asymptotic behavior of the energy on the repulsive side for large $\gamma$, given in~\cite{Batchelor_2006}. The gray dashed line on the right is the asymptotic value $s(\infty)=0.749\dots$ of the dimensionless contact density at infinite repulsion.}
\label{fig:paircorrfct-gs-1d-balanced-plot}
\end{figure}
%%%End: Figure of the g2 plot and s plot
%%%%%%%%%%%%%%%%%%%%%%%%%%%%%%%%%%%%%%%%%%%%%%%%%%%%%%%%%

In figure \ref{fig:paircorrfct-gs-1d-balanced-plot}, we show the pair distribution function as a function of the dimensionless coupling $\gamma$ for both the attractive and
the repulsive gas as determined from a numerical solution of the Gaudin-Yang integral equations. 
Despite superficial appearance, the function is non-analytic at $\gamma=0$ because the presence of
pairing on the attractive side $\gamma<0$ leads to an expansion of the form ($\theta(-\gamma)$ is the usual theta-function)
\begin{equation}
 \label{eq:attractive}
g_{\uparrow \downarrow}^{(2)}(0, \gamma<0)=1+|\gamma|\theta(-\gamma) -\frac{\gamma}{\pi^2}\ln^2(|\gamma|)+\ldots \, .
\end{equation} 
The contribution $|\gamma|\theta(-\gamma)$ arises from the two-body binding energy and dominates the behavior in the 
limit $|\gamma|\gg 1$. The logarithmic term, in turn, is a non-analytic correction to the mean field energy in the presence of 
pairing, whose many-body spin-gap $\Delta\sim\exp{-\pi^2/2|\gamma|}$ is exponentially small as $\gamma\to 0^{-}$~\cite{Krivnov_1975, Fuchs_2004}. 
On the repulsive side, the probability for two fermions with opposite spin to be at the same point in space continuously decreases to zero with increasing strength of the repulsion. In the limit $\gamma\gg 1$ the decay is $\sim 1/\gamma^2$. As we will see, this reflects the fact that the contact density approaches a finite, universal value for a Fermi gas with infinite repulsion.
To discuss the tail in the momentum distribution, it is convenient to define a dimensionless measure $s$ of the value of the contact density $\mathcal{C}=s\cdot k_F^4$ by factoring out the Fermi wave vector $k_F=\pi n/2$ as the characteristic scale on which the (intensive) momentum distribution $n_{\sigma}(k)$ varies. For the ground state of the balanced gas, equation \eqref{eq:contact-translational-invariant} then implies the simple connection
 \begin{equation}
\label{eq:s-definition}
 s=\frac{4}{\pi^4}\gamma^2\cdot g_{\uparrow\downarrow}^{(2)}(0)=\frac{8}{\pi^4}\gamma^2 e'(\gamma)\, ,
\end{equation}
between the dimensionless contact density $s$ and the derivative of the ground state energy $e(\gamma)$.  It is shown
quantitatively in figure \ref{fig:paircorrfct-gs-1d-balanced-plot}, again for both positive and negative values of $\gamma$.  For a weakly repulsive (or also attractive) gas,
$s(\gamma\to 0)=4\gamma^2/\pi^4$ vanishes as it should, because the local pair distribution function 
$g_{\uparrow\downarrow}^{(2)}(0)$ approaches one in this limit.  The gas with infinite repulsion, in turn, has a finite value 
\begin{equation}
\label{eq:s-Fermi}
s(\infty)=32\ln{2}/(3\pi^2)\simeq 0.7491252
\end{equation}
of the dimensionless contact density, which follows from the asymptotic behavior $e(\gamma)=e(\infty)(1-4\ln{2}/\gamma +\ldots)$ of the ground state energy ~\cite{Batchelor_2006}.  In fact, a similar result also applies to bosons in one dimension with an infinite zero range 
repulsion. This is a Tonks-Girardeau gas  ~\cite{Bloch_2008_review}, whose equation of state and pair distribution function coincides with
that of a free Fermi gas with Fermi wave vector $k_F=\pi n_B$. 
Its momentum distribution, however, is very different. At small momenta, it diverges like $1/\sqrt{k}$ 
while for large momenta it exhibits a $\mathcal{C}_B/k^4$-tail as shown by Olshanii and Dunjko ~\cite{Olshanii_2003_1D}.
Using equation \eqref{eq:bosonic-contact}, which relates the bosonic contact to the pair correlation function at zero distance, 
the asymptotic behavior $g^{(2)}(0)=4 \pi^2 /3 \gamma_B^2$ in the limit $\gamma_B\gg 1$ ~\cite{Gangardt_2003} gives 
rise to a universal value 
\begin{equation}
\label{eq:s-Bose}
s_B(\infty)=\frac{4}{3\pi^2}\simeq 0.1350949
\end{equation}
of the dimensionless contact $s_B=\mathcal{C}_B/k_F^4$, which is much smaller than that for infinitely repulsive fermions.
The finiteness of the dimensionless contact $s(\gamma)$ in the limit of infinite repulsion can be inferred from a simple argument.
Indeed, at $\gamma=\infty$, the momentum distribution $n_{\sigma}(k)=f(x)\leq 1$ must be a universal function of the dimensionless
variable $x=k/k_F$ with a power law decay $s/x^4$ at $x\gg 1$. The normalization $\int \! dx f(x)=1$ together with the 
fact that the power law $s/x^4$ appears for $x=\mathcal{O}(1)$ due to the absence of another momentum scale in the problem
beyond $k_F$ 
then implies that $s(\infty)$ must be of order one.  Using equation  \eqref{eq:s-definition}, the finiteness of $s(\infty)$ immediately
implies that the energy $e(\gamma)$ approaches a universal constant $e(\infty)$ in the limit of infinite repulsion.  As will be shown
below, a completely analogous line of arguments applies to the 3D repulsive Fermi gas with contact interactions. \\

\begin{figure}
 \begin{center}
 \includegraphics[width=0.8\columnwidth]{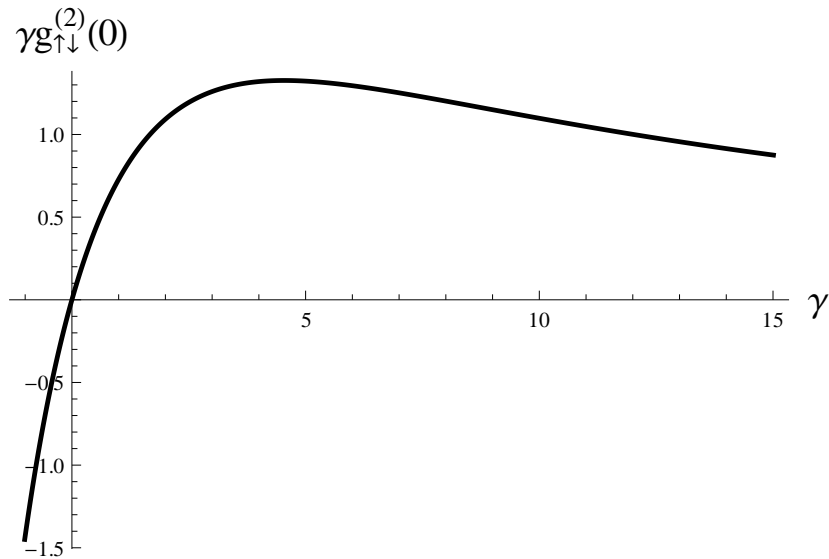}
 \end{center}
\caption{Plot of the pair-correlation function $g_{\uparrow\downarrow}^{(2)}(0)$ times $\gamma$, determining the prefactor of the $1/q^2$-tail of the static structure factor in equation \eqref{eq:structurefactor-asymptotic}}
\label{fig:gammag2-groundstate-plot}
\end{figure}

Apart from the pair distribution at vanishing distance and the related dimensionless contact density $s$ that are shown in figure \ref{fig:paircorrfct-gs-1d-balanced-plot}, the homogeneous balanced gas is also characterized by the combination 
$a_1\mathcal{C}(a_1)\sim \gamma g_{\uparrow\downarrow}^{(2)}(0)$ that appears both in the pressure relation 
\eqref{eq:pressure-relation} and in the asymptotics  \eqref{eq:structurefactor-asymptotic} of the static structure factor. 
This combination is essentially the interaction energy per particle. For the ground state of the balanced homogeneous gas,
it can be written in the simple form
\begin{equation}
 \label{eq:interaction-energy}
\frac{\langle H'\rangle}{N}=\frac{\hbar^2 n^2}{2m}\cdot \gamma e'(\gamma)
\end{equation}   
where we have used the equations \eqref{eq:Hprime} and \eqref{eq:contact-translational-invariant}. A plot of the product 
$\gamma g_{\uparrow\downarrow}^{(2)}(0)$ is shown in figure \ref{fig:gammag2-groundstate-plot}.
%%%%%%%%%%%%%%%%%%%%%%%%%%%%%%%%%%%%%%%%%%%%%%%%%%%%%%%%%
%%%Begin: Figure of the gamma * g2 plot
%%%End: Figure of the gamma * g2 plot
%%%%%%%%%%%%%%%%%%%%%%%%%%%%%%%%%%%%%%%%%%%%%%%%%%%%%%%%%
It exhibits a maximum at $\gamma=4.541\ldots$ with value
$1.3266\ldots$. Physically, this maximum comes about because the interaction energy of the two-component Fermi gas
becomes small both as $\gamma\ll 1$ and for $\gamma\gg 1$. In the latter limit,  a strong zero range repulsion is equivalent 
to having effectively a Pauli exclusion principle also between fermions of opposite spin. The total energy is thus essentially
of kinetic origin. The fact that $e(\infty)=4e(0)$ shows that the gas with infinite repulsion has the same ground state energy
as a single-component gas with twice the density, where the associated doubling of the Fermi wave vector
increases the energy by a factor four. In the limit of infinite repulsion, therefore, the balanced gas with total spin zero is
energetically degenerate with a fully polarized ferromagnetic ground state. Note, however, that despite the fact that both states have 
the same energy and, in fact, have equations of state $p=2\mathcal{E}$ that are identical at any temperature, their momentum 
distribution functions are quite different. In the fully polarized state it is the step function of an ideal Fermi gas with an 
effective $\tilde{k}_F=2k_F=\pi n$. The unpolarized state, in turn, is a Luttinger liquid with a tail $\mathcal{C}/k^4$ that has
a universal finite value of the contact density $\mathcal{C}(\infty)\simeq0.749 k_F^4$. The fact that ferromagnetism in one 
dimension only appears at infinite coupling is consistent with the Lieb--Mattis theorem~\cite{Lieb_1962} . In the following
section, we will show that this result can be derived by a simple variational 
argument, even in the absence of the explicit Bethe Ansatz solution for the ground state energy. 

%%%%%%%%%%%%%%%%%%%%%%%%%%%%%%%%%%%%%%%%%%%%%%%%%%%%%%%%%%%%%
%%Absence of saturated Ferromagnetism for zero range repulsive 
\section{Absence of Ferromagnetism for zero range interactions} \label{sec:ferromagnetism}
The problem of deriving either necessary or sufficient conditions for the appearance of 
a ferromagnetic ground state of itinerant fermions is a major and - to a large extent - 
unsolved problem in many-body physics. A classic result in this context is the theorem of 
Lieb and Mattis that one never gets ferromagnetism in the ground state 
of itinerant fermions in 1D with spin-independent interactions~\cite{Lieb_1962, Aizenman_1990}. 
The Fermi gas with two-body contact interactions discussed here is just a special case of
this and correspondingly cannot have a ferromagnetic ground state. As we have seen above,
however, the ground state of this system in the infinite repulsion limit becomes degenerate
with the fully polarized ferromagnet. As was shown by Lieb and Mattis,  an infinite coupling constant 
is indeed a {\it necessary} condition for the appearance of saturated ferromagnetism in one dimension. 
This result is a simple consequence of a variational upper bound 
on the ground state energy but does {\it not} require the explicit solution of the many-body problem by 
the Bethe Ansatz.  Remarkably, as will be shown below, a completely analogous argument also 
applies in three dimensions, ruling out saturated ferromagnetism for 
repulsive fermions at finite values of $k_Fa$, with $a>0$ the scattering length of the 
pseudopotential interaction. This may be viewed as a continuum, low-density analog of a result 
proven by Kollar, Strack and Vollhardt ~\cite{Strack_1994, Strack_1996} that there is no saturated ferromagnetism
in the half-filled fermionic Hubbard model with finite on-site repulsion $U$ unless one adds further
ferromagnetic exchange interactions, e.g. to nearest neighbors.  

For a balanced, homogeneous 1D Fermi gas, the Tan adiabatic theorem in the form
\eqref{eq:s-definition} together with the fact that the contact density $s(\gamma)$ remains finite at infinite coupling, 
quite generally implies that the dimensionless energy $e(\gamma)$ approaches a finite value as 
$\gamma\to\infty$. In the absence of the Bethe Ansatz solution, the Tan relations alone, however,  
give no information about the asymptotic value $e(\infty)$. In particular, it is 
impossible to decide on the basis of the Tan relations alone whether the ground state energy 
$e^{(\rm{F})}=4 e(\gamma=0)=\pi^2/3$ of a saturated ferromagnet might be below that of the
normal Luttinger liquid state with zero polarization beyond a {\it finite} critical value of the repulsive
interaction.  A simple variational argument, however, shows that this cannot happen, in accordance
with the Lieb--Mattis theorem. To see this we note that, for contact interactions, the fully polarized 
ferromagnet    
\begin{equation}
\label{eq:Ferromagnet}
\langle \{ z_l \} \vert \Psi^{(\rm{F})} \rangle = \rm{det}\,\phi_l \left( \{z_l\} \right) \, \Ket{ S=N/2}
\end{equation}
is a Slater determinant of the $N$ lowest single-particle states $\phi_l$ for $l=1,\ldots N$ combined 
with the symmetric state $\Ket{ S=N/2}$ with maximal spin (note that $N=N_{\uparrow}+N_{\downarrow}$ 
is even). Following a well known argument which shows that
for just two fermions with spin-independent interactions the ground state is always a singlet~\cite{Lieb_1962},
we now choose a variational state of the form 
\begin{equation}
\label{eq:Slater}
\langle \{ z_l \} \vert \Psi_{\rm{var}} \rangle = \left| \rm{det} \,\phi_l \left( \{z_l\} \right) \right| \, \Ket{ S=0}\, .
\end{equation}
It describes an unpolarized Fermi gas with vanishing magnetization which combines the symmetric spatial wave-function 
formed by taking the absolute value of a Slater determinant in \eqref{eq:Ferromagnet} with the completely antisymmetric 
spin state $\Ket{ S=0 }$. Its spatial wave-function  
describes a gas of $N$ {\it bosons} with an infinite zero range repulsion. In one dimension, 
this is a Tonks-Girardeau gas, whose energy is known to be equal to that of a single-component, non-interacting 
Fermi gas with wave vector $\pi n$~\cite{Girardeau_1960, Lieb_Liniger_1963_Bose}.  
The variational state $\Ket{\Psi_{\rm var}}$ therefore provides
an upper bound $e(\gamma)\leq \pi^2/3$ on the dimensionless ground state energy of
the unpolarized, repulsive gas. Since $\Ket{\Psi_{\rm var}}$ has vanishing probability for any of the fermions 
to be at the same point in space, it can be (and actually {\it is}) an exact eigenstate only at
$\gamma=\infty$, where $g_{\uparrow\downarrow}^{(2)}(0)\equiv 0$. The limit $e(\infty)=\pi^2/3$
of degeneracy with a fully polarized ferromagnet can therefore only be reached at a point,
where the two-component Fermi gas completely fermionizes in the 
sense that there is an effective Pauli principle even for particles with opposite spin.    
Without an explicit solution of the problem, the argument thus rules out saturated 
ferromagnetism in 1D at any finite coupling. 

The argument above is instructive but of course it gives no really new information because the ground
state of 1D fermions with contact interactions is known exactly by the Bethe Ansatz and ferromagnetism of any
kind - not just the simple fully saturated ferromagnet considered here -  is excluded for much more general 
interactions by the Lieb-Mattis theorem. Remarkably, however, the argument can be extended in a straightforward 
manner to the case of fermions in 3D which interact with a repulsive pseudopotential. In this case, 
due to the existence of a two-body bound state for positive scattering lengths, the true equilibrium state is  
an s-wave superfluid that evolves from the nontrivial unitary gas at $k_Fa=\infty$ toward an ideal Bose-Einstein
condensate of dimers in the limit $k_Fa\to 0^+$~\cite{Ketterle_2008_review, Bloch_2008_review, Giorgini_2008_review}. 
The repulsive normal Fermi liquid at positive scattering lengths therefore corresponds to a metastable branch of the 
3D balanced Fermi gas with contact interactions. In the experiment~\cite{Jo_2009_ferro}, the gas is initially prepared 
in the weakly interacting limit $k_Fa\ll 1$ and then the interactions are quickly ramped to the strongly repulsive regime
$k_Fa\gg 1$ in order to avoid production of bound dimers~\cite{Zwerger_2009_ferro}. An analysis of the 
dynamic competition between a putative formation of ferromagnetic domains within an RPA-approximation 
and the formation of pairs indicates that the growth rate of the latter instability is always faster~\cite{Pekker_2010_RPA}.
%(for a discussion of the dynamic aspects of the formation of ferromagnetic domains in the re 
%geometry see ~\cite{Conduit_2009}). 
In particular, as shown by these authors, anomalies of the kind seen experimentally can be explained in terms
of pair formation rather than incipient ferromagnetism (see also~\cite{Ho_2011}).

In the following, we will ignore the issue of the dynamics and discuss the question whether a Stoner instability 
might occur in the strongly repulsive Fermi 3D gas with contact interactions, assuming a decay to bound pairs
happens on a time scale much longer than the equilibration time of the repulsive gas (thus, effectively, the 
state with bound pairs is projected out).  Since the Tan relations are based on operator identities, they also apply to 
the repulsive branch of the gas. This branch is characterized by a dimensionless contact density 
$s_r={\mathcal{C}}_r/k_F^4$ which is determined by the asymptotic decay of the momentum distribution 
(the subscript in $s_r$ refers to repulsive branch). 
Defining a dimensionless energy $e(k_Fa)$ per particle, normalized such that $e(0)=1$, via
\begin{equation}
 \label{eq:groundstate-energy_3D}
\frac{E_r}{N}=\frac{3}{5}\frac{\hbar^2 (3\pi^2n)^{2/3}}{2m}\cdot e(k_Fa)
\end{equation} 
in a manner analogous to equation \eqref{eq:groundstate-energy} in the 1D case,
the Tan adiabatic theorem in 3D~\cite{Tan_2008_adiabatic, Braaten_2010_Book} reads
\begin{equation}
\label{eq:adiabatic3D}
 \frac{d\, e(k_Fa)}{d(k_Fa)}=\frac{5\pi}{2(k_Fa)^2}\cdot s_r(k_Fa)\, .
\end{equation}
%The contact $s(v)$ of the repulsive gas is a positive and monotonically decreasing function of $v$. 
From the standard textbook result for the dimensionless energy $e(k_Fa)$ of the repulsive Fermi gas to 
second order in $k_Fa$~\cite{Abrikosov_1963}, the contact density in the weak coupling limit $k_Fa\ll 1$ is given by
\begin{equation}
\label{eq:contact3D}
s_r(k_Fa\ll 1)=\frac{4}{9\pi^2}(k_Fa)^2+\frac{16\left(11-2\ln(2)\right)}{105 \pi^3}(k_Fa)^3+\ldots \, ,
\end{equation}
where the leading term just accounts for the mean field contribution. 
Similar to the argument above in 1D, the Tan adiabatic theorem  \eqref{eq:adiabatic3D} in 3D implies that the energy $e(k_Fa)$ 
necessarily approaches a finite constant $e(\infty)$ in the infinite repulsion limit because the dimensionless
contact $s_r$ of the repulsive gas cannot diverge. 
In order to prove or disprove the appearance of saturated ferromagnetism of the repulsive 3D gas with contact interactions, 
it is necessary to know whether the constant $e(\infty)$ that characterizes the infinite repulsion limit of the dimensionless energy
is above or below the value $e^{(F)}=2^{2/3}=1.5874...$ reached for a fully 
polarized (and thus ideal!) Fermi gas with Fermi wave vector $\tilde{k}_F=2^{1/3}k_F$. The existence of a 
ferromagnetic  instability is suggested by a renormalized Hartree-Fock
calculation, where a finite magnetization in the ground state appears in a discontinuous manner for $k_Fa\geq 1.054$ 
and full polarization is reached already for $k_Fa\geq 1.112$~\cite{Duine_2005}.  These values are not far from what is 
obtained by extending the perturbative result
\begin{equation}
\label{eq:Landau}
F_0^a(k_Fa)=-\frac{2}{\pi}k_F a -\frac{8}{3\pi^2}\left(1-\ln(2)\right)\left(k_Fa\right)^2+\ldots \, ,
\end{equation}
for the Landau parameter that determines the spin susceptibility\footnote{Recent measurements of the spin susceptibility of the 
unitary Fermi gas in its normal state have been performed by Sommer {\it{et al.}}~\cite{Zwierlein_2011}.} 
\begin{equation}
\label{eq:chi_s}
\chi_s\sim\,\frac{m^{\star}/m}{1+F_0^a}
\end{equation}
of a Fermi liquid to $k_Fa$ values of order one, predicting a divergent $\chi_s$ with $F_0^a\vert_c=-1$ at a finite 
critical value $k_Fa\vert_c\simeq \pi/2+\ldots$. 
Moreover, the appearance of {\it saturated} ferromagnetism in the ground state of the repulsive gas is supported by 
Monte Carlo calculations~\cite{Bertaini_2010_ferro, Trivedi_2010_ferro} which find that the energy of the 
unpolarized repulsive gas rises beyond that of a fully polarized state at a critical coupling $k_Fa\vert_c$ of order one. 
An extension of the variational argument used above in the 1D case shows, however, that this cannot happen   
for pure contact interactions. 
To see this, we choose as a variational state for the 3D repulsive Fermi gas with vanishing magnetization a state of 
the form given in \eqref{eq:Slater}. This state has no off-diagonal long range order and is thus orthogonal to the superfluid 
ground state of the 3D pseudopotential at positive scattering length. Its spatial wave-function describes a gas of bosons 
at infinite repulsion $n_B^{1/3}a=+\infty$. Such a gas is known to be effectively fermionized, i.e. its energy per particle is of order 
$\hbar^2n^{2/3}/m$~\cite{Cowell_2002}. The tendency to fermionization in 3D is, however, less pronounced in 3D than in 1D.
Specifically, as found by Song and Zhou within a Bogoliubov type Ansatz for the bosonic formulation of the problem, 
the energy of the 3D Bose gas with infinite repulsion stays below that of a single-component Fermi gas at the same total density 
$n=n_B$ by about $20$ percent~\cite{Zhou_2009}. Thus, there is an upper bound $e(\infty)\leq 0.8\, e^{(F)}$ on the energy 
of the repulsive, unpolarized Fermi gas which excludes the appearance of a fully saturated ferromagnet at finite values of $k_Fa$. 
An important point to note in this context is that the argument is restricted to zero range interactions. Moreover, it rules out 
only saturated ferromagnetism for the {\it homogeneous} gas not, however, partially polarized states or a non-vanishing
spin in the ground state of a few-body system.  Indeed, the ground state of three or five fermions in an isotropic harmonic 
trap has non-zero angular momentum $l=1$~\cite{Nishida_2007_CFT}.  As a result, three 
fermions with the same spin can be put in the three degenerate sublevels of the $l=1$ manifold, thus giving
rise to a fully polarized ground state of three fermions in a trap, as noted by Liu et. al.~\cite{Drummond_2010}. The underlying degeneracy of the ground state, however, does not extend to larger particle numbers and, indeed, our argument above rules out saturated magnetism as the ground state of the homogeneous gas. In practice, for ultracold gases, the repulsive branch at positive scattering lengths is only metastable. Therefore, a calculation of the energy on this branch alone is not sufficient to determine the conditions for ferromagnetism at least as a metastable state \cite{Pekker_2010_RPA}. For example, a variational calculation of the energy of a single flipped spin in the background of  a fully polarized up-spin Fermi sea with wave vector $k_{F \uparrow}$ indicates that the fully polarized state is stable if $k_{F \uparrow} a > 2.35$ \cite{Zhai_10}. This calculation, which is analogous to that of the spin-flip energy in the Nagaoka state of the lattice Hubbard model discussed below, does not account for the finite lifetime of the repulsive polaron. The latter has recently been determined by a functional renormalization group approach \cite{Schmidt_Enss_2011}. It turns out that in the 'ferromagnetic' regime $k_{F \uparrow} a > 1.57$, the lifetime $ \hbar / \Gamma $ is smaller than about $5 \hbar / \epsilon_{F \uparrow}$. The fully polarized state therefore decays on a microscopic timescale that rules out fully saturated ferromagnetism even as a metastable state.  

Within a broader context, the conclusion that a single band model with zero range interactions shows
no saturated ferromagnetism at finite values of the coupling constant is in fact not surprising. Indeed,  
related questions have been studied in considerable detail in the context of the 'nearly ferromagnetic' normal phase of
Helium 3, where $F_0^a\simeq -0.7$ at atmospheric pressure. As discussed by Vollhardt~\cite{Vollhardt_1984},
under compression this system does not become ferromagnetic with $F_0^a$ approaching $-1$.  Instead, it is the 
effective mass $m^{\star}$ in equation \eqref{eq:chi_s} that diverges, signaling a transition to a solid. Helium 3 
is therefore an almost localized rather than a nearly ferromagnetic Fermi liquid~\cite{Vollhardt_1984}. Of course,
the instability to a solid phase requires a hard core part of the interaction, which is not present in 
the case of ultracold atoms. Therefore, it might appear that for pure contact interactions a ferromagnetic
ground state is not pre-empted by other instabilities. Approximate calculations of the Fermi liquid parameters 
for the relevant simple Stoner Hamiltonian with an interaction of the form $U\int n_{\uparrow}(x)n_{\downarrow}(x)$
however indicate that $F_0^a>-0.63$ is bounded from below for arbitrary large values of the repulsion 
$U$~\cite{Ainsworth_1983}, thus excluding a ferromagnetic instability in this 
simple model. This is consistent with an earlier argument by Kanamori, which states that the effective repulsion $U^{*}$ 
that enters the Stoner criterion remains finite even in the limit of a diverging bare repulsion $U$~\cite{Kanamori_1963}.
The stability of saturated ferromagnetic ground states has been studied extensively also in lattice models
of the Hubbard type~\cite{Gebhard}. In this case, saturated ferromagnetism appears for infinite on-site repulsion 
near half filling, because for a single hole it is energetically favorable to move in a background of fully aligned 
spins~\cite{Nagaoka_1966}. On the basis of variational wave-functions it is possible to show, however, 
that the associated saturated ferromagnetism is restricted to the regime near half filling but does not survive 
at low densities, where continuum physics applies. For instance, in a simple cubic lattice 
the Nagaoka state is only stable for hole dopings $\delta\leq 0.24$~\cite{Hanisch_1997}. 
%Quite generally,  in agreement with our argument above, the tendency towards a saturated 
%ferromagnetic ground state becomes weaker with increasing dimensionality~\cite{Hanisch_1997}. 
A single band Stoner model with zero range repulsion is therefore not adequate to account
for ferromagnetism, which - in practice - requires the presence of flat, degenerate bands and exchange 
interactions of finite range~\cite{Strack_1996}.

%%%%%%%%%%%%%%%%%%%%%%%%%%%
%Conclusion
\section{Conclusion} \label{sec:conclusion}
In this work, we have extended the Tan relations for fermions with contact interactions
to the situation in one dimension.  We have calculated the associated Tan 
contact in the ground state for arbitrary interaction strengths from the exact Bethe-Ansatz
results for the energy per particle. In particular, we have shown that there is a universal 
constant in the asymptotic decay of the momentum distribution in the limit of 
very strong repulsion which is connected with the saturation of the ground state energy
as a function of the coupling constant. As in the closely related 3D case, the Tan relations are 
associated with the short range behavior of the one- and two-particle density matrix. 
As such, they may be viewed as complementing the universal features that characterize
quantum liquids in one dimension in terms of their long-distance, low-energy 
behavior as Luttinger liquids~\cite{Haldane_1981, Giamarchi}.  
An important feature of the Tan relations is their applicability to 
an arbitrary state of the system. The assumption of contact interactions, in turn, appears as
a strong restriction. As emphasized by Braaten~\cite{Braaten_2010_Book} and earlier by 
Zhang and Leggett~\cite{Leggett_2009}, however,
the Tan relations hold more generally, provided the interaction range $r_0$ is much smaller 
than any other characteristic length scale, in particular the average interparticle spacing
$n^{-1}$ and the thermal wavelength $\lambda_T$. 

A surprising conclusion that appears from our analysis is that a combination of the Tan relations 
with a variational argument rules out saturated
ferromagnetism in repulsive Fermi gases with contact interactions not only in the - well known - 
case of one dimension but also in three dimensions. In essence, the argument is based on
the observation that fully saturated ferromagnetism for zero range interactions requires 
fermionization even of particles with opposite spin. This is reached asymptotically for 
infinite repulsion in 1D. In the 3D situation, however, the tendency to fermionization is less
pronounced. Correspondingly, no saturated ferromagnetism appears in this case at 
finite values of the repulsion. Of course, it remains an open
problem to study this issue in the presence of a finite range interaction, where both the 
Stoner instability and the transition to a solid ground state are possible
in 3D ~\cite{Trivedi_2010_ferro}. Moreover, in view of the well known Kohn--Luttinger 
argument~\cite{Kohn_1965}, the ground state of the repulsive gas is likely to be a (p-wave) superfluid,
whose transition temperature might become of order $T_F$ if $k_Fa\gg 1$. 

The possibility to realize strongly interacting Fermi gases with tunable interactions in one 
dimension with ultracold gases in optical lattices ~\cite{Moritz_2005, Bloch_2008_review}
%{\it the paper is H. Moritz et al., Phys. Rev. Lett. 94, 210401 (2005)}
opens the chance to experimentally test the relations derived here. This is of particular interest since 
they apply even in situations that are still far from the ground state which is hard to reach in one dimension
due to the typically long equilibration times.     

\section*{Acknowledgements}

This work was started when one of the authors (W.Z.) was a visitor at the Physics Department 
of the University of Trento. The hospitality of the members of the cold atom group there and a number of
discussions with - in particular - Stefano Giorgini, Lev Pitaevskii, Alessio Recati and Sandro Stringari are
gratefully acknowledged. The authors also would like to thank Eugene Demler, Tilman Enss, Florian Gebhard, Wolfgang Ketterle, Matthias Punk, Mohit Randeria,  
Richard Schmidt, Kurt Sch\"onhammer, Boris Spivak and Martin Zwierlein for useful discussions. 
The work was completed at the Institute for Nuclear Theory at the University of Seattle in the
program on 'Fermions from Cold Atoms to Neutron Stars: Benchmarking the Many-Body Problem'. It is a pleasure
to acknowledge support through this program and discussions with Jason Ho, Dam Thanh Son, Shizhong Zhang and Fei Zhou. 
Part of ths work has been supported by the DFG research unit ``Strong Correlations in Multiflavor Ultracold Quantum Gases''.

%%%%%%%%%%%%%%%%%%%%%%%%%%%%%%%
%%begin:Appendix
\begin{appendix}
 \section{Feynman rules} \label{app:feynman}
 The matrix elements needed to establish the Operator Product Expansions in sections \ref{sec:asymptotics} and \ref{sec:dd-correlator} via matching can be calculated using the Feynman rules for the theory defined by the Hamiltonian density in equation \eqref{eq:hamilton-density}. They are supplemented by the rules for the operator vertices, which can be derived by explicitly Wick-contracting. Alternatively, in an approach using standard quantum mechanical scattering theory, one could also think of the operator vertices as matrix elements of the operators in the non-interacting theory. Note that in this approach, the sums of the diagrams in sections \ref{sec:asymptotics} and \ref{sec:dd-correlator} are the analogues to the full matrix element $\bra{\pm p'} \Omega_{-}^{\dagger} A  \Omega_{+} \ket{\pm p}$ of some operator $A$, where $\Omega_{\pm}$ are the M\o ller operators from scattering theory, such that this a matrix element between scattering states that have \textit{in}-asymptote $\ket{\pm p}$ and \textit{out}-asymptote $\ket{\pm p'}$.

\subsection{Standard Diagrams}
We start with the standard Feynman rules of the theory, needed to evaluate, for example, the diagrams leading to the Amplitude $\mathcal{A}$ given in \eqref{eq:scattering-amplitude}.

\paragraph{Momenta}
Assign a two-momentum $(p_0,p)$, where $p_0$ denotes the energy and $p$ the momentum, to all external lines. The two-momenta of internal lines are constrained by conservation of energy and momentum at each interaction vertex.

\begin{figure}[b]

\begin{center}
\subfigure[{$\frac{i}{q_0 - \frac{q^2}{2m_{\uparrow}} + i\epsilon}$}]{\label{subfig:feyn1} \includegraphics[width=60pt,height=2pt]{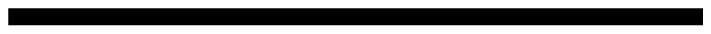}} \hspace{5pt}
\subfigure[{$\frac{i}{q_0 - \frac{q^2}{2m_{\downarrow}} + i\epsilon}$}]{\label{subfig:feyn2} \includegraphics[width=60pt,height=2pt]{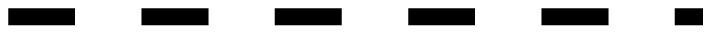}} \hspace{5pt}
\subfigure[{$-ig_1$}]{\label{subfig:feyn3} \includegraphics[width=50pt,height=75pt]{fig1_part2_scattering_amplitude_naked.eps}} \hspace{5pt}
\end{center}

%caption: basic Feynman diagrams
\caption{Basic diagrams for the Feynman rules}
\label{fig:feynmanrules}
\end{figure}

\paragraph{Propagators}
For each internal line of species $\sigma$ with two-momentum $(q_0,q)$, assign a propagator factor $i/(q_0 - q^2 / 2m_{\sigma} + i \epsilon )$. Diagrammatically, the propagation of $\uparrow$-particles is symbolized by a solid line (see figure \ref{subfig:feyn1}), the propagation of $\downarrow$-particles by a dashed line (see figure \ref{subfig:feyn2}).

%begin: picture of some basic Feynman diagrams
%end: picture of some basic Feynman diagrams

\paragraph{Vertices}
For each interaction vertex (see figure \ref{subfig:feyn3}), assign a factor $-ig_1$. Conserve energy and momentum at each such vertex.

\paragraph{Loop momenta}
Integrate over all two-momenta $(q_0,q)$, which are not determined by the two-momenta of the external lines and two-momentum conservation at the interaction vertices, with the measure $\int dq_0 dq / (2\pi)^2$.

\subsection{Operator vertices}
The various operators occurring during the matching contribute their part to the matrix elements as well. We list their contributions to the diagrams involving \textit{two-particle scattering} in order of their appearance. In particular, the operator vertices do not need to conserve energy or momentum. Let $q$ and $-q$ be the momenta entering, where the $\uparrow$-particle carries momentum $q$ and the $\downarrow$-particle momentum $-q$. Let further $q'$ and $-q'$ be the momenta leaving, such that the $\uparrow$-particle carries $q'$ and the $\downarrow$-particle carries $-q'$. We state explicitly, that the one-particle operators all have in common delta functions for the momenta of the particles they don not act on. This delta-function is of course no longer there in diagrams only containing one particle, such as those in figure \ref{fig:localbilocalop-matrixelements-1pstate}. Further, for the rules describing one-particle operators of the $\downarrow$-species, just revert the signs of the momenta.

\begin{center}
\renewcommand\arraystretch{2.0} %Spacing between Table rows = 2 times standard value
\begin{tabular}{lr}
\textbullet \, $\psi_{\uparrow}^{\dagger} (R) \psi_{\uparrow} (R+x)$: & $e^{i(q-q')R} e^{iqx} \delta(-q+q')$ \\ 
\textbullet \, $\psi_{\uparrow}^{\dagger} ( \partial^{m}_{x} \psi_{\uparrow} ) (R)$: & $(iq)^m e^{i(q-q')R} \delta(-q+q')$ \\ 
\textbullet \, $\psi_{\uparrow}^{\dagger} \psi_{\downarrow}^{\dagger} \psi_{\downarrow} \psi_{\uparrow} (R)$: & $1$ \\ 
\textbullet \, $\psi_{\uparrow}^{\dagger} \psi_{\uparrow} (R-\frac{x}{2}) \psi_{\downarrow}^{\dagger} \psi_{\downarrow} (R+\frac{x}{2})$: & $e^{-i(q-q')x}$ \\ 
\textbullet \, $\psi_{\uparrow}^{\dagger} \psi_{\uparrow} ( \partial_x \psi_{\downarrow}^{\dagger} \psi_{\downarrow}) (R)$ & $-i(q-q')$ \\ 
\textbullet \, $( \partial_x \psi_{\uparrow}^{\dagger} \psi_\uparrow) \psi_{\downarrow}^{\dagger} \psi_{\downarrow} (R)$: & $i(q-q')$
\end{tabular}
\renewcommand\arraystretch{1.0} %Restore standard value
\end{center}

%\begin{widetext}
\subsection{Example diagram}
Let us evaluate the diagram in figure \ref{subfig:1pdm4} with incoming momenta $\pm p$ and outgoing momenta $\pm p'$, which is the part of the one-particle density matrix element that produces the non-analyticities, and thus plays a crucial role for the study of the tail of the momentum distribution. Using the Feynman rules given above, the diagram is given as the integral
\begin{equation}
\label{eq:na-me-1pdm-startingterm}
 i \mathcal{A}(E) i \mathcal{A}(E') \int \frac{dq_0 dq}{(2\pi)^2} e^{iqx} \frac{i}{E - q_0 - \frac{q^2}{2m_{\uparrow}} + i\epsilon} \frac{i}{E' - q_0 - \frac{q^2}{2m_{\uparrow}} + i\epsilon} \frac{i}{q_0 - \frac{q^2}{2m_{\downarrow}} + i\epsilon} ,
\end{equation}
where we again used the shorthands $E=p^2/2m_r$, $E'=p'^2/2m_r$. The $q_0$-integration can be carried out using the Residue theorem. This removes one propagator factor and contributes a factor of $(-i)$, leading to
\begin{equation}
\label{eq:na-me-1pdm-afterq0}
\begin{split}
- i \mathcal{A}(E) i \mathcal{A}(E') \int \frac{dq}{2 \pi} e^{iqx} \frac{1}{E - \frac{q^2}{2m_{r}} + i\epsilon}\frac{1}{E' - \frac{q^2}{2m_{r}} + i\epsilon} \\ = 4 m_r^2 \mathcal{A}(E) \mathcal{A}(E') \frac{i}{2} \frac{1}{|p|^2-|p'|^2} \left[  \frac{e^{i|p||x|}}{|p|} - \frac{e^{i|p'||x|}}{|p'|} \right],
\end{split}
\end{equation}
where we have used the Residue theorem again, and carried out the limit $\epsilon \rightarrow 0$. For $x>0$, one has to close the contour through the upper complex plane, for $x<0$ one has to use the lower one.
%\end{widetext}
\end{appendix}
%end:Appendix
%%%%%%%%%%%%%%%%%%%%%%%%%%%%%%%

%%%%%%%%%%%%%%%%%%%%%%%%%%%%%%%
% bibliography
\bibliographystyle{model1a-num-names}
\bibliography{tanrelations_1d.bib}
%
%%%%%%%%%%%%%%%%%%%%%%%%%%%%%%%
\end{document}